\documentclass[preprint,12pt]{elsarticle}

\usepackage{amsmath,amssymb,amsfonts}
\usepackage{graphicx}
\usepackage{xcolor}
\usepackage{tikz}
\usetikzlibrary{arrows.meta,positioning}
\usepackage[colorlinks=true,allcolors=blue]{hyperref}
\newcommand{\rev}[1]{#1}
\newcommand{\revB}[1]{#1}

\newcommand{\Rei}{\mathfrak{R}_i}
\newcommand{\Imi}{\mathfrak{I}_i}
\newcommand{\Rej}{\mathfrak{R}_j}
\newcommand{\Imj}{\mathfrak{I}_j}

\journal{Wave Motion}
\begin{document}
\begin{frontmatter}

\title{Wave scattering around a submerged vertical permeable breakwater\tnoteref{t1}}
\tnotetext[t1]{Accepted manuscript of the article published in \emph{Wave Motion}, \url{https://doi.org/10.1016/j.wavemoti.2026.103854}. \copyright~2026. This manuscript version is made available under the CC-BY-NC-ND 4.0 license \url{https://creativecommons.org/licenses/by-nc-nd/4.0/}.}

\author[gt]{Jeongin Kim}
\author[snu,icee]{Yong Sung Park\corref{cor1}}
\cortext[cor1]{Corresponding author.}
\ead{dryspark@snu.ac.kr}
\affiliation[gt]{organization={School of Civil and Environmental Engineering, Georgia Institute of Technology},
  city={Atlanta}, state={GA}, country={United States of America}}
\affiliation[snu]{organization={Department of Civil, Urban and Environmental Engineering, Seoul National University},
  city={Seoul}, country={Republic of Korea}}
\affiliation[icee]{organization={Institute of Construction and Environmental Engineering, Seoul National University},
  city={Seoul}, country={Republic of Korea}}

\begin{abstract}
An analytical solution for a wave velocity field scattered by a submerged permeable vertical plate-type breakwater under the linear monochromatic wave is obtained and the applications of the solution are presented.
The water has an infinite depth, and the flow is assumed to be incompressible, inviscid, and irrotational, which leads to the two-dimensional potential wave theory. The permeable breakwater vertically occupies a finite interval beneath the water surface and the water flows through the breakwater. \revB{The porous boundary condition is linear, but it couples the normal velocity on the plate to the discontinuity in the velocity potential across it, so that the scattering problem does not reduce directly to a Riemann--Hilbert problem of standard form. This is resolved by a perturbation expansion in a small dimensionless parameter representing the permeability. The solution is expanded up to the first order: the leading order gives the wave scattered by the impermeable breakwater through a homogeneous Riemann--Hilbert problem, and the first-order term gives the correction for the permeability through a nonhomogeneous Riemann--Hilbert problem. The far-field amplitudes follow in closed form up to a single Cauchy principal-value integral.} \rev{The effects} of wavelength, breakwater length, and breakwater permeability conditions on the reflection and transmission coefficients are discussed in detail as an illustrative example of the application of the solution.
\rev{An exact energy identity is also derived; it verifies the first-order solution and yields a closed-form \revB{energy-consistency bound} $\varepsilon_{\max}(kb)$ \revB{on the perturbation parameter, beyond which the truncated expansion ceases to conserve energy}.} \revB{The hydrodynamic load on the plate is obtained to the same order, and the leading-order results are verified against the closed-form solution of Evans for the impermeable plate.}
\end{abstract}


\end{frontmatter}

\section{Introduction}
Waves from deep water propagate into the shoreline and affect the coastal area, often causing severe problems. Therefore, various shapes and functions of breakwaters have been investigated and constructed in maritime and offshore areas to dissipate wave energy from the open sea.
In recent decades, there has been a growing body of research that studies plate-type vertical breakwaters. For example, Briggs et al. [1] argued that near-surface partial barriers are particularly effective for controlling waves in deep water, and Chwang \& Chan [2] maintained that porous barriers can dissipate more wave energy and are more suited to reducing wave forces on the barriers.

\revB{Analytical and semi-analytical treatments of wave scattering by vertical porous barriers are by now numerous, and the relevant ones are reviewed below. The great majority, however, concern barriers that are surface-piercing or bottom-standing, or barriers in water of finite depth treated by eigenfunction expansions or integral equations solved numerically. A closed-form solution for a submerged porous plate of finite extent, both of whose edges lie in the fluid, has to the best of our knowledge not been given. The present paper addresses this case.}

For a solid vertical breakwater, Dean [3] first solved the wave scattering problem due to the semi-infinite vertical barrier at a distance below the water surface. Ursell [4] obtained the velocity field of waves by a finite thin plate oscillating with a small angle, following the work of [5]. Subsequently, several authors, notably Lewin [6] and Mei [7], have contributed to generalizing the problem of wave generation and scattering by an arbitrary number of vertical breakwaters in deep water. Later, Evans [8] solved the diffraction problems on a completely submerged rolling plate in closed form, determining the velocity potential everywhere in the fluid. Wave scattering in finite water by vertical barrier was solved based on Galerkin approximation in [9].

Macaskill [10] first numerically solved the water wave reflection by a permeable barrier, considering the permeable barrier as an impermeable barrier with numerous gaps. Chwang [11] developed a porous-wavemaker theory and showed that the porosity can reduce the hydrodynamic force impinging on the wavemakers as well as the wave amplitude. This theory was used in investigating porous breakwater by Lee \& Chwang [12]. Liu \& Li [13] used velocity potential decomposition in the breakwater to find a solution for wave reflection and transmission by a surface-piercing porous breakwater. Gayen \& Mondal [14] used a hypersingular integral equation for the discontinuity of the velocity potential across the plate and numerically solved the equation with a Chebyshev polynomial. Manam \& Sivanesan [15] proposed an analytical approach to find the scattering of deep water waves by a bottom-standing submerged or a surface-piercing vertical porous barrier. \rev{Their approach connects the porous-barrier problem to the corresponding solid-barrier problems and yields explicit reflection coefficients for arbitrary porosity; it relies, however, on the barrier and the gap together partitioning the vertical axis (a surface-piercing barrier, or a submerged barrier extending downwards indefinitely), and therefore does not cover the plate of finite extent considered here, both of whose edges lie in the fluid.} Using the eigenfunction method and a coupled boundary element-finite difference method, Koley \& Sahoo [16] found the wave scattering by permeable vertical flexible membrane barriers. Sasmal et al. [17] studied the oblique wave diffraction and the energy dissipation due to two unequal vertical porous wave barriers using Galerkin approximation. \revB{More recent work has extended porous-barrier scattering in several directions. Sarkar et al. treated multiple thin porous walls and two non-uniform permeable walls in water of finite depth [18, 19], a pair of porous barriers beside a bottom-standing rectangular obstacle by Havelock's expansion and Galerkin's approximation [20], and, by a multi-term Galerkin method, a submerged L-shaped Jarlan-type breakwater beneath an elastic cover [21]; Behera et al. [22] studied wave trapping by a porous barrier near a rigid wall over a step-type bottom by the eigenfunction expansion method with a multi-mode approximation of the modified mild-slope equation, and Paul \& Behera [23] analysed wave attenuation on a floating rigid dock by an array of surface-piercing perforated barriers with a multi-domain boundary element method; Sahoo et al. examined, by the eigenfunction expansion method, a floating elastic plate with double bottom-standing porous structures in front of a seawall [24], and, by a hybrid boundary element--finite difference method, the scattering of capillary-gravity waves by surface-piercing porous barriers in the presence of a uniform current over a porous sea bed [25]. These studies considerably extend the configurations treated, and they also make the methodological contrast clear: all of them resolve the porous condition numerically for an arbitrary porous-effect parameter, mostly in water of finite depth, whereas the present work seeks an explicit reduction in the small-permeability limit for a plate of finite extent in infinite depth.}

\revB{In the present study, an analytical solution for the velocity potential of the scattering problem around a submerged vertical permeable plate of finite extent in deep water is obtained. The model assumptions are those of linear potential theory in water of infinite depth, with the through-flow across the plate governed by a Darcy-type resistance law, so that the porosity parameter is real and small. The method differs from the eigenfunction-expansion, Galerkin and integral-equation approaches cited above in that the permeable-plate condition is expanded in the small porosity parameter $\varepsilon$, and each order is then reduced, by the method of reduced potentials, to a Riemann--Hilbert problem that is solved in closed form. The leading order recovers the impermeable-plate solution of Evans [8], and the first order gives the permeability correction as an explicit integral representation, in which a single finite Hilbert transform remains to be evaluated numerically.

The contributions of this work, to the best of our knowledge, are the following. (i) The first-order permeability correction for a submerged plate of finite extent is obtained in explicit form; the analytical approach of Manam \& Sivanesan [15] does not reach this geometry. (ii) An exact energy identity is derived for the full permeable-plate problem, relating the reflected, transmitted and dissipated energy fluxes at any porosity. (iii) Its first-order form yields a nontrivial identity between the far-field amplitude of the first-order solution and a weighted norm of the leading-order potential jump on the plate, which the derived solution is shown to satisfy identically and which therefore serves as an independent check on the analysis. (iv) The truncated expansion is shown to conserve energy if and only if $\varepsilon$ lies below a bound $\varepsilon_{\max}(kb)$ that is obtained in closed form, and within this bound the dissipated energy flux $1 - R^2 - T^2$ is predicted in closed form at first order.}

The paper is organized as follows. First, the governing equation and the boundary conditions are \revB{formulated} in \S2. The solution of the leading-order problem and its application are presented in \S3. Subsequently, the correction effect of the first-order problem will be analyzed in \S4. \rev{The energy identity and the \revB{energy-consistency range} of the perturbation expansion are derived in \S5.} \revB{The numerical evaluation and the results, including the loads on the plate, are presented in \S6.} Finally, conclusions will be discussed in \S\rev{7}.

\section{Formulation}
\subsection{Statement of problem}
We consider a problem involving the interaction of linear monochromatic water waves and a thin flat permeable plate as illustrated in figure 1. Here, we take the Cartesian coordinate system $(x, y)$ where the $x$-axis is the mean free surface, and the $y$-axis \revB{points} vertically upwards. Under the water surface, a permeable plate occupies a finite interval $L : x = 0,\ -b < y < -a$. On the water surface, a train of waves with the small amplitude $A$ and the angular wave frequency $\omega$ propagates from $x = +\infty$ in the negative $x$-direction.

\begin{figure}[htbp]
\centering
\begin{tikzpicture}[scale=1.1]
  \draw[->] (-3.2,0) -- (3.4,0) node[below] {$x$};
  \draw[->] (0,-2.6) -- (0,0.9) node[right] {$y$};
  \draw[thick] plot[domain=-3.1:3.2, samples=120] (\x, {0.08*sin(6*\x r)});
  \draw[line width=2.4pt] (0,-0.7) -- (0,-2.1);
  \foreach \yy in {-0.8,-1.0,...,-2.0} { \draw[white, line width=0.9pt] (-0.045,\yy) -- (0.045,\yy); }
  \node[left] at (-0.08,-0.7) {$-a$};
  \node[left] at (-0.08,-2.1) {$-b$};
\end{tikzpicture}
\caption{A vertical plate occupies a finite interval along the $y$ axis and a regular wave train of small amplitude propagates along the $x$-axis. \rev{(The authors' original diagram (2023), redrawn with Anthropic Claude (Fable~5).)}}
\end{figure}
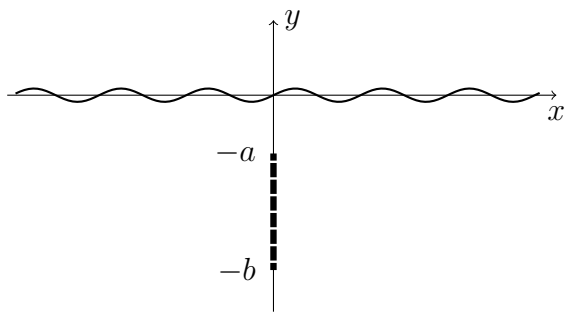

In order to solve the problem in closed form using potential wave theory, it is assumed that the fluid has infinite depth and is incompressible, inviscid, and irrotational. Also, we assume that the wave amplitude is small and there is no wave breaking. With the first assumption, the viscous effects on the boundary layer of the fluid can be neglected since we are interested in the wave scattering problem away from the plate. \rev{Two quantitative restrictions accompany the small-amplitude assumption. First, the formulation is the leading order of a Stokes expansion in the wave steepness, so the neglected free-surface nonlinearities enter at $O(kA)$ and the results below apply for $kA \ll 1$. Second, the plate must remain submerged throughout the wave cycle, which requires the wave amplitude to be smaller than the submergence depth of the upper edge, $A < a$; for the smallest ratio considered below ($a/b = 0.001$), this condition confines the results to correspondingly small amplitudes; indeed, since $A < a$ is more restrictive than any fixed steepness criterion $kA < s$ whenever $ka < s$, it becomes the binding restriction before wave breaking does in the limit $a/b \to 0$ at fixed $kb$.}

For an irrotational flow in \rev{two dimensions}, there exists a velocity potential $\Phi(x, y, t)$, and the \rev{gradient of the potential field yields the velocity field, $\mathbf{v} = (u, v) = \nabla\Phi$, which is the convention consistent with the linearized Bernoulli equation (2.13) below}. This velocity potential will also have the frequency of $\omega$ and be harmonic in time. Without loss of generality, the velocity potential can be written as,
\begin{equation}
\Phi(x, y, t) = \Rej\left\{ \phi(x, y)e^{-j\omega t} \right\}, \tag{2.1}
\end{equation}
where $\phi(x, y)$ is the spatial velocity potential, and $j$ is the time-related imaginary unit defined as $j = \sqrt{-1}$.

Taking the two-dimensional linearized potential wave theory, the governing equation and the combined free surface boundary condition are,
\begin{equation}
\frac{\partial^2\phi}{\partial x^2} + \frac{\partial^2\phi}{\partial y^2} = 0, \qquad \text{in the fluid}, \tag{2.2}
\end{equation}
\begin{equation}
k\phi - \frac{\partial\phi}{\partial y} = 0, \qquad \text{on } y = 0. \tag{2.3}
\end{equation}
Here, $k = \omega^2/g$ is the wavenumber in the deep water.

Since we are considering \rev{infinitely} deep water, the fluid velocity will vanish as $y \to -\infty$. Thus,
\begin{equation}
\nabla\phi \to 0, \qquad y \to -\infty. \tag{2.4}
\end{equation}

The linearity of the formulation allows us to split the potential into the incident and scattered wave potential. The incident wave velocity potential $\Phi_I$ is also represented \revB{by} the combination of spatial-related potential and the time-harmonic part:
\begin{equation}
\Phi_I(x, y, t) = \Rej\left\{ \phi_I(x, y)e^{-j\omega t} \right\}. \tag{2.5}
\end{equation}
Given the water surface elevation due to the incident wave as
\begin{equation}
\eta_I(x, t) = -\frac{1}{g}\frac{\partial \Phi_I|_{y=0}}{\partial t} = A\cos(kx + \omega t), \tag{2.6}
\end{equation}
the incident spatial velocity potential in equation (2.5) would be
\begin{equation}
\phi_I(x, y) = -\frac{jgA}{\omega}e^{ky - jkx}. \tag{2.7}
\end{equation}

The scattered wave propagating to the positive side of the $x$-axis will be superposed with the incident wave, and the scattered wave propagating to the negative side of the $x$-axis will travel outwards to $-\infty$. Therefore,
\begin{align}
\phi^{+\infty}(x, y) &\sim X^{+\infty}e^{ky+jkx} - \frac{jgA}{\omega}e^{ky-jkx}, & x \to +\infty, \tag{2.8}\\
\phi^{-\infty}(x, y) &\sim X^{-\infty}e^{ky-jkx}, & x \to -\infty, \tag{2.9}
\end{align}
for some constants $X^{\pm\infty}$. Superscripts $(\cdot)^{+\infty}$ and $(\cdot)^{-\infty}$ mean $(\cdot)$ at $x \to \pm\infty$, respectively. $X^{+\infty}e^{ky+jkx}$ in (2.8) represents the spatial velocity potential generated by the reflected wave propagating to $+x$ direction, and $X^{-\infty}e^{ky-jkx}$ in (2.9) represents the spatial velocity potential due to the transmitted wave propagating to $-x$ direction.

The velocity is bounded everywhere, but at the edges of the plate, the velocity may be unbounded and permit a mild, integrable singularity as below:
\begin{equation}
\frac{\partial\phi}{\partial r} = O\left(\frac{1}{r^{\lambda}}\right), \quad 0 < \lambda < 1, \text{ near } x = 0,\ y = -a, -b. \tag{2.10}
\end{equation}
Here, $r$ is the distance from the edges of the plate. \revB{The exponent $\lambda$ is determined by the solution. The Riemann--Hilbert analysis of \S3 selects $\lambda = 1/2$ (see the remark following (3.8)), so that the velocity has the inverse-square-root singularity expected at the sharp edges of a thin plate [8, 26], while the singularity remains integrable and the energy finite.}

\subsection{Permeable boundary condition on the plates}
As suggested by Taylor [27], the boundary condition on the permeable breakwater is obtained with the assumption that the flow through the permeable plate is due to the pressure difference between both sides of the plate.
\begin{equation}
\frac{\partial\Phi}{\partial x} = -\frac{B}{\rho\nu}(p^+ - p^-), \qquad \text{on } L. \tag{2.11}
\end{equation}
Here, $\rho$ is the density of water, $\nu$ is the kinematic viscosity of the water, and $p$ is the pressure applied \revB{to} the plate. Superscripts $(\cdot)^+$ and $(\cdot)^-$ mean $(\cdot)$ on the positive side and the negative side of the plates, respectively. $B$ represents the permeability of the plate having a dimension of length [11]. In this paper, we defined $B$ as
\begin{equation}
B = \frac{\kappa}{D}, \tag{2.12}
\end{equation}
where $\kappa$ is the permeability of the plate and $D$ is the thickness of the plate. \revB{Relation (2.12) follows from Darcy's law: the seepage velocity through a thin porous layer of thickness $D$ driven by the pressure difference $p^+ - p^-$ across it is $u = \kappa\,(p^+ - p^-)/(\mu D)$, with $\mu = \rho\nu$ the dynamic viscosity [28], and comparison with (2.11) gives $B = \kappa/D$, which has the dimension of length as required of the constant introduced by Taylor [27] and Chwang [11].}

Meanwhile, for an infinitesimal amplitude, Bernoulli's equation may be linearized, and the pressure can be expressed as
\begin{equation}
p(x, y, t) = -\rho\frac{\partial\Phi}{\partial t}(x, y, t) - \rho g y. \tag{2.13}
\end{equation}
Substituting (2.13) into (2.11),
\begin{equation}
\frac{\partial\Phi}{\partial x} = \frac{B}{\nu}\left(\frac{\partial\Phi^+}{\partial t} - \frac{\partial\Phi^-}{\partial t}\right), \qquad \text{on } L. \tag{2.14}
\end{equation}
Since $\Phi(x, y, t) = \phi(x, y)e^{-j\omega t}$ from (2.1), we can simplify (2.14) as below:
\begin{equation}
\frac{\partial\phi}{\partial x} = -j\frac{B\omega}{\nu}(\phi^+ - \phi^-), \qquad \text{on } L. \tag{2.15}
\end{equation}
\revB{Condition (2.15) is linear in $\phi^+$ and $\phi^-$, but it prescribes the normal velocity on the plate in terms of the unknown discontinuity of the potential across it, so that the two faces of the plate are coupled. For an impermeable plate the condition $\partial\phi/\partial x = 0$ makes the real part of the reduced potential introduced in \S3 vanish on $L$, and this is precisely what reduces the problem to a Riemann--Hilbert problem of standard form. Under (2.15) the normal velocity is instead proportional to the potential discontinuity, so that this reduction is not available. The perturbation method of \S2.3 decouples the condition order by order, and the Riemann--Hilbert problems are then solved in sequence.}

\subsection{Perturbation expansion}
\revB{Define the dimensionless permeability parameter $\varepsilon$ as}
\begin{equation}
\varepsilon = \frac{B\omega}{\nu k}. \tag{2.16}
\end{equation}
\rev{The parameter $\varepsilon$ is of the same form as the porosity parameter of the porous-wavemaker theory [11, 12], and it measures the ratio of the seepage velocity driven through the plate by the pressure difference to the orbital velocity of the wave; $\varepsilon \ll 1$ therefore corresponds to a hydraulically nearly-solid, fine-pored plate. Assuming that the breakwater is composed of pervious materials, $\kappa$ \revB{is of order} $10^{-7}$--$10^{-10}\ \mathrm{m}^2$ [28, p.~136]. For other parameters, $D \sim O(1\ \mathrm{m})$ and $\nu \sim O(10^{-6}\ \mathrm{m}^2/\mathrm{s})$, while $\omega$ and $k$ are tied by the deep-water dispersion relation $k = \omega^2/g$, so that $\omega \sim O(1\ \mathrm{s}^{-1})$ implies $k \sim O(10^{-1}\ \mathrm{m}^{-1})$; consequently $\varepsilon$ ranges over $O(10^{-3}-1)$ for materials of practical interest. It should be noted, however, that the expansion below is asymptotic in $\varepsilon\to 0$, and the precise range of $\varepsilon$ in which the truncated expansion remains energy-consistent is derived in closed form in \S5; the upper end of the practical range may fall outside it, particularly for short waves.}

Expanding $\phi$ into a perturbation series of the form,
\begin{equation}
\phi = \phi_0 + \varepsilon\phi_1 + \cdots, \tag{2.17}
\end{equation}
in which $\phi_0$ represents the spatial velocity potential due to the impermeable plates, and $\phi_1$ shows the correction when the plate has permeability. Substituting (2.17) into (2.15), we can get the boundary conditions of the leading-order and the first-order spatial velocity potential functions, respectively, along the vertical plate:
\begin{align}
\frac{\partial\phi_0}{\partial x} &= 0, \tag{2.18}\\
\frac{\partial\phi_1}{\partial x} &= -jk(\phi_0^+ - \phi_0^-). \tag{2.19}
\end{align}
\revB{The expansion below is an expansion about the impermeable-breakwater problem rather than a tool for linearization. The expansion (2.17) is a regular perturbation in $\varepsilon$, since the small parameter enters only as the coefficient of the right-hand side of (2.15), and it multiplies neither the highest derivative in the governing equation nor that in any boundary condition, so that the leading-order problem retains the full structure of the original one [29]. Each order is governed by the same Laplace equation, free-surface condition and radiation conditions, and by the plate conditions (2.18)--(2.19), in which the permeability appears only through a forcing determined by the previous order; this is the structure of the small-porosity limit of the porous-wavemaker theory of Chwang [11]. Because (2.18) and (2.19) prescribe the normal velocity on the plate, each order admits the reduction to a Riemann--Hilbert problem employed in \S\S3--4. The expansion is asymptotic as $\varepsilon \to 0$; the range of $\varepsilon$ over which the truncated expansion remains energy-consistent is determined in closed form in \S5, where the extension to a complex-valued porosity parameter is also discussed.}

Also, in view of (2.8) and (2.9), we require the behavior of $\phi_0$ at $x \to \pm\infty$ to be,
\begin{align}
\phi_0^{+\infty}(x, y) &\sim X_0^{+\infty}e^{ky+jkx} - \frac{jgA}{\omega}e^{ky-jkx}, & x \to +\infty, \tag{2.20}\\
\phi_0^{-\infty}(x, y) &\sim X_0^{-\infty}e^{ky-jkx}, & x \to -\infty, \tag{2.21}
\end{align}
and for $\phi_1$,
\begin{align}
\phi_1^{+\infty}(x, y) &\sim X_1^{+\infty}e^{ky+jkx}, & x \to +\infty, \tag{2.22}\\
\phi_1^{-\infty}(x, y) &\sim X_1^{-\infty}e^{ky-jkx}, & x \to -\infty. \tag{2.23}
\end{align}
\revB{The solution procedure is summarised in figure 3.}

\section{Leading-order solution}
\subsection{Leading-order velocity potential}
Define the complex potential $w_0(z)$ as,
\begin{equation}
w_0(z) = \phi_0(x, y) + i\psi_0(x, y), \qquad \Imi\{z\} < 0, \tag{3.1}
\end{equation}
where $z = x + iy$, $i = \sqrt{-1}$ and $\psi_0(x, y)$ is the two-dimensional stream function. \rev{Here $i$ is the imaginary unit of the spatial complex plane $z=x+iy$, whereas $j$ in (2.1) is the imaginary unit of the temporal phasor; although both satisfy $i^2=j^2=-1$, they act on different complexifications and are kept separate throughout: $\Rei$, $\Imi$ denote the real and imaginary parts with respect to $i$, and $\Rej$, $\Imj$ those with respect to $j$. A quantity such as $\phi$ may thus be $i$-real while remaining $j$-complex.}

In addition, since the breakwater has the form of a thin plate, it is convenient to consider the reduced potential $W_0(z)$, defined by,
\begin{equation}
W_0(z) = \frac{dw_0}{dz} + ikw_0, \qquad \Imi\{z\} < 0. \tag{3.2}
\end{equation}
By the use of the reduced potential, from (2.3),
\begin{equation}
\Imi\{W_0(z)\} = 0, \quad \text{along the horizontal axis } (y=0). \tag{3.3}
\end{equation}
If the imaginary part of $W_0(z)$ is zero on \revB{the} $x$-axis, for $\Imi\{z\} > 0$, $W_0(z)$ can be continued by Schwarz's reflection principle into $y > 0$, where,
\begin{equation}
W_0(z) = \overline{W_0(\bar{z})}. \tag{3.4}
\end{equation}
Since $W_0(z)$ is a single-valued function outside the circle $|z| = b$, assuming that (2.8) and (2.9) may be differentiated once with respect to $x$ or $y$, it follows that,
\begin{equation}
W_0(z) = O(1), \qquad |z| \to \infty, \tag{3.5}
\end{equation}
which means that $W_0(z)$ is bounded as $z$ goes to infinity. In addition, from (2.18), the real part of $W_0(z)$ is zero on the plate, and can be expanded through the whole complex plane using (3.4):
\begin{equation}
\Rei\{W_0(z)\} = 0, \qquad \text{on } L + L', \tag{3.6}
\end{equation}
where $L'$ is the interval $x = 0,\ a < y < b$, the reflection of $L$ with respect to the real axis. Finally, $W_0(z)$ may be unbounded near the ends of $L$, $L'$. Thus,
\begin{equation}
W_0(z) = O\left(\frac{1}{r^{\lambda}}\right), \quad 0 < \lambda < 1, \text{ near } z = \pm ia, \pm ib. \tag{3.7}
\end{equation}
The problem of determining $W_0(z)$ satisfying (3.5), (3.6), and (3.7) is a typical homogeneous Riemann-Hilbert problem.

The solution of the homogeneous Riemann-Hilbert problem for the plane with a cut along a straight line is given by Muskhelishvili [26, p. 261]:
\begin{equation}
W_0(z) = \frac{C_0 + D_0 z^2}{\sqrt{(z^2 + a^2)(z^2 + b^2)}}. \tag{3.8}
\end{equation}
Here, $C_0$ and $D_0$ are real constants with respect to $i$. \rev{In the classification of [26], this is the solution unbounded at all four endpoints (index $\kappa = 2$), which fixes the edge exponent in (2.10) and (3.7) to $\lambda = 1/2$; the same inverse-square-root behaviour carries over to the first-order solution (4.15).} \revB{The value $\lambda = 1/2$ thus follows from (3.8) and is not an assumption. Condition (3.7) requires only that the singularity be integrable, and the solution of the homogeneous problem under this requirement is (3.8), whose behaviour near each endpoint is $O(r^{-1/2})$.} The complex potential $w_0(z)$ is found by integrating (3.8):
\begin{equation}
w_0(z) = e^{-ikz}\left[B_0 + \int_{-ia}^{z} e^{ik\zeta}W_0(\zeta)\, d\zeta\right], \tag{3.9}
\end{equation}
where $B_0$ is an arbitrary real constant with respect to $i$.

It remains to determine the constants $B_0$, $C_0$, and $D_0$. First, the radiation boundary conditions prescribed in (2.20) and (2.21) will be imposed onto (3.9).

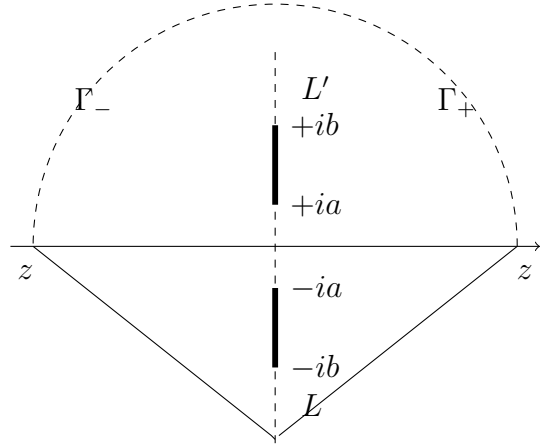
\begin{figure}[htbp]
\centering
\begin{tikzpicture}[scale=1.0]
  \draw[->] (-3.5,0) -- (3.5,0);
  \draw[dashed] (0,-2.6) -- (0,2.6);
  \draw[dashed] (0,0) ++(0:3.2) arc (0:180:3.2);
  \node at (-2.4,1.9) {$\Gamma_-$};
  \node at (2.4,1.9) {$\Gamma_+$};
  \draw[line width=2.2pt] (0,-0.55) -- (0,-1.6);
  \draw[line width=2.2pt] (0,0.55) -- (0,1.6);
  \node[right] at (0.06,0.55) {$+ia$};
  \node[right] at (0.06,1.6) {$+ib$};
  \node[right] at (0.06,-0.55) {$-ia$};
  \node[right] at (0.06,-1.6) {$-ib$};
  \node[right] at (0.2,2.1) {$L'$};
  \node[right] at (0.2,-2.1) {$L$};
  \node[below] at (-3.3,-0.1) {$z$};
  \node[below] at (3.3,-0.1) {$z$};
  \draw (-3.2,0) -- (-0.05,-2.5) -- (0,-2.55);
  \draw (3.2,0) -- (0.05,-2.5);
\end{tikzpicture}
\caption{Integration path for $w_0(z)$ as $z \to \pm\infty$. \rev{(The authors' original diagram (2023), redrawn with Anthropic Claude (Fable~5).)}}
\end{figure}

\begin{figure}[htbp]
\centering
\revB{\begin{tikzpicture}[node distance=5.5mm, every node/.style={font=\small},
  box/.style={draw, rounded corners=2pt, align=center, text width=0.86\textwidth, inner sep=4pt},
  arr/.style={-{Latex[length=2mm]}, thick}]
\node[box] (b1) {\textbf{Formulation} (\S2): linear potential theory, infinite depth; permeable-plate condition (2.15) couples $\partial\phi/\partial x$ to $\phi^+-\phi^-$ on $L$};
\node[box, below=of b1] (b2) {\textbf{Regular expansion in the porosity parameter} $\varepsilon = B\omega/(\nu k)$ (2.16)--(2.17): $\partial\phi_0/\partial x = 0$, $\ \partial\phi_1/\partial x = -jk(\phi_0^+-\phi_0^-)$ on $L$};
\node[box, below=of b2] (b3) {\textbf{Reduced potentials} $W_n = dw_n/dz + ikw_n$ (3.2), (4.2): real part prescribed on $L+L'$, bounded at infinity, integrable edge singularity};
\node[box, below=of b3] (b4) {\textbf{Riemann--Hilbert problems}: homogeneous (3.8) at leading order, nonhomogeneous (4.15) at first order; constants fixed by the far-field and zero-circulation conditions (3.22)--(3.30), (4.36)--(4.40)};
\node[box, below=of b4] (b5) {\textbf{Far-field amplitudes}: $R_0,T_0$ in closed form (3.45)--(3.46); first-order correction $\Lambda$ (4.47) with one finite Hilbert transform $F(u)$ (4.26), evaluated by Chebyshev interpolation (\S6)};
\node[box, below=of b5] (b6) {\textbf{Checks and outputs}: energy identity (5.1), first-order identity (5.4), energy-consistency bound $\varepsilon_{\max}$ (5.6); $R$, $T$, dissipation, plate loads (\S\S5--6)};
\foreach \i/\j in {b1/b2, b2/b3, b3/b4, b4/b5, b5/b6} \draw[arr] (\i) -- (\j);
\end{tikzpicture}}
\caption{\revB{Outline of the solution procedure.}}
\end{figure}

For $z \to +\infty$, let us modify the integral term in (3.9) as,
\begin{equation}
\begin{split}
\int_{-ia}^{z} e^{ik\zeta}W_0(\zeta)\, d\zeta = \int_{-ia}^{0^+ + i\infty} e^{ik\zeta}W_0(\zeta)\, d\zeta &+ \int_{i\infty+0^+}^{z} e^{ik\zeta}(W_0(\zeta) - D_0)\, d\zeta\\
&- \frac{iD_0}{k}e^{ikz}.
\end{split} \tag{3.10}
\end{equation}
When the second integral in (3.10) is taken following a large arc, which is \revB{denoted} $\Gamma_+$ in figure 2, this term vanishes by Jordan's lemma, since $W_0(z)$ is bounded at infinity and $W_0(z) - D_0 \to 0$ as $|z| \to \infty$. Thus, for $z \to +\infty$ the complex potential reduces to,
\begin{equation}
w_0^{+\infty}(z) \sim e^{-ikz}\left[B_0 + \int_{-ia}^{i\infty+0^+} e^{ik\zeta}W_0(\zeta)\, d\zeta\right] - \frac{iD_0}{k}. \tag{3.11}
\end{equation}
Contracting the integral term in (3.11) onto the $y$-axis from the right, the integration should be separately examined for some interval parts on account of the branch cut $(a, b)$ in the reduced potential. To simplify the expression, it is convenient to define the following functions representing each interval along the integration path,
\begin{align}
a_1(k) &= \int_a^b \frac{e^{-ku}}{\sqrt{(u^2 - a^2)(b^2 - u^2)}}\, du, \tag{3.12}\\
a_2(k) &= \int_{-a}^{a} \frac{e^{-ku}}{\sqrt{(a^2 - u^2)(b^2 - u^2)}}\, du, \tag{3.13}\\
a_3(k) &= \int_b^{\infty} \frac{e^{-ku}}{\sqrt{(u^2 - a^2)(u^2 - b^2)}}\, du, \tag{3.14}
\end{align}
and their second derivative with respect to $k$,
\begin{equation}
a_i''(k) = \frac{d^2 a_i}{dk^2}, \qquad i = 1, 2, 3. \tag{3.15}
\end{equation}
Then, for $z \to +\infty$,
\begin{equation}
\begin{split}
w_0^{+\infty}(z) \sim e^{-ikz}\big[&B_0 - \big(C_0 a_1(k) - D_0 a_1''(k)\big)\\
&+ i\big(\big(C_0 a_2(k) - D_0 a_2''(k)\big) - \big(C_0 a_3(k) - D_0 a_3''(k)\big)\big)\big] - \frac{iD_0}{k}.
\end{split}\tag{3.16}
\end{equation}
Similarly, for $z \to -\infty$, the integration path of an integral part in (3.9) is taken as a union of a vertical upward line contracted to the left side of the $y$-axis and an arc \revB{denoted} $\Gamma_-$. Because of the branch cut on $(a, b)$, the imaginary part of $W_0^-(iu)$ has a different sign from $W_0^+(iu)$. For the sake of brevity, the following functions are defined for each interval:
\begin{align}
\gamma_0(k) &= C_0 a_1(k) - D_0 a_1''(k), \tag{3.17}\\
\alpha_0(k) &= C_0 a_2(k) - D_0 a_2''(k), \tag{3.18}\\
\beta_0(k) &= C_0 a_3(k) - D_0 a_3''(k). \tag{3.19}
\end{align}
Thus, the asymptotic behavior of leading-order complex potential as $z \to \pm\infty$ is expressed in terms of $\gamma_0(k)$, $\alpha_0(k)$, and $\beta_0(k)$:
\begin{equation}
w_0^{\pm\infty}(z) \sim e^{-ikz}\left[B_0 \mp \gamma_0(k) + i(\alpha_0(k) - \beta_0(k))\right] - \frac{iD_0}{k}. \tag{3.20}
\end{equation}
Taking the real part of (3.20) with respect to $i$, the leading-order spatial velocity potential for $z \to \pm\infty$ can be obtained:
\begin{equation}
\phi_0^{\pm\infty}(x, y) \sim e^{ky}\left[(B_0 \mp \gamma_0(k))\cos kx + (\alpha_0(k) - \beta_0(k))\sin kx\right]. \tag{3.21}
\end{equation}
Substituting this into the radiation boundary conditions (2.20) and (2.21), the value of $B_0$ is found as,
\begin{equation}
B_0 = -\frac{jgA}{\omega}, \tag{3.22}
\end{equation}
and $X_0^{\pm\infty}$ as,
\begin{align}
X_0^{+\infty} &= -\gamma_0(k) = -j\left(\alpha_0(k) - \beta_0(k) + \frac{gA}{\omega}\right), \tag{3.23}\\
X_0^{-\infty} &= \gamma_0(k) - \frac{jgA}{\omega} = j(\alpha_0(k) - \beta_0(k)). \tag{3.24}
\end{align}
The relation of $\alpha_0(k)$, $\beta_0(k)$, and $\gamma_0(k)$ in (3.23) and (3.24) can be rearranged with respect to $C_0$ and $D_0$, and this gives one equation in terms of $C_0$ and $D_0$:
\begin{equation}
(a_1(k) - j(a_2(k) - a_3(k)))C_0 - \left(a_1''(k) - j\left(a_2''(k) - a_3''(k)\right)\right)D_0 = \frac{jgA}{\omega}. \tag{3.25}
\end{equation}
In addition to the radiation boundary condition, the zero-circulation condition is applied to find $C_0$ and $D_0$.
\begin{equation}
\Rei\left\{\oint_{\Gamma} e^{ik\zeta}W_0(\zeta)\, d\zeta\right\} = 0. \tag{3.26}
\end{equation}
Assuming that the circulation around a plate is zero and applying the zero-circulation condition (3.26) with contracting the path of integration onto $L$, we have,
\begin{equation}
\Rei\left\{\oint_{\Gamma} e^{ik\zeta}W_0(\zeta)\, d\zeta\right\} = \int_a^b \frac{e^{ku}(C_0 - D_0 u^2)}{\sqrt{(u^2 - a^2)(b^2 - u^2)}}\, du = 0. \tag{3.27}
\end{equation}
Introducing the functions in (3.12) and (3.17), a simplified expression of (3.27) is given below:
\begin{equation}
\gamma_0(-k) = a_1(-k)C_0 - a_1''(-k)D_0 = 0. \tag{3.28}
\end{equation}
Thus, combining (3.25) and (3.28), the unknown constants $C_0$ and $D_0$ can be determined after some algebra:
\begin{align}
C_0 &= \frac{jgA}{\omega}\frac{a_1''(-k)}{\Delta_{123}}, \tag{3.29}\\
D_0 &= \frac{jgA}{\omega}\frac{a_1(-k)}{\Delta_{123}}, \tag{3.30}
\end{align}
where,
\begin{align}
\Delta_{123} &= \Delta_{11} - j(\Delta_{12} - \Delta_{13}), \tag{3.31}\\
\Delta_{1i} &= \begin{vmatrix} a_i(k) & a_1(-k)\\ a_i''(k) & a_1''(-k) \end{vmatrix}, \qquad \text{for } i = 1, 2, 3. \tag{3.32}
\end{align}
Thus, the leading-order term of the complex potential is fully determined below:
\begin{equation}
w_0(z) = -\frac{jgA}{\omega}e^{-ikz}\left[1 - \frac{1}{\Delta_{123}}\int_{-ia}^{z} \frac{e^{ik\zeta}\left(a_1''(-k) + a_1(-k)\zeta^2\right)}{\sqrt{(\zeta^2 + a^2)(\zeta^2 + b^2)}}\, d\zeta\right]. \tag{3.33}
\end{equation}
Taking the real part of (3.33) with respect to $i$ and multiplying the harmonic term $e^{-j\omega t}$, the leading-order velocity potential is obtained.

\subsection{Wave scattering by an impermeable plate}
An analytical solution can be utilized in a wide variety of ways. \revB{One use of the solution is to evaluate} the wave attenuation efficiency by comparing the reflected and transmitted wave amplitude and that of the incident wave. Especially for the leading-order solution, the reflected and transmitted waves indicate the waves scattered by an impermeable plate.

Let the leading-order reflection coefficient $R_0$ and the leading-order transmission coefficient $T_0$ \rev{be} defined as the ratio of the wave amplitude of the leading-order component of the reflected and transmitted wave, respectively, at $x \to \pm\infty$ to the incident wave amplitude:
\begin{equation}
R_0 = \frac{A_{R0}}{A}, \quad T_0 = \frac{A_{T0}}{A}, \tag{3.34}
\end{equation}
where $A_{R0}$ and $A_{T0}$ \revB{denote} the leading-order reflected and transmitted wave amplitude at $x \to \pm\infty$, respectively.

Substituting the coefficients for the leading-order spatial velocity potential derived as in (3.23), and (3.24), the leading-order spatial velocity potentials as $x \to \pm\infty$ (2.20) and (2.21) are rephrased as,
\begin{align}
\phi_0^{+\infty}(x, y) &\sim -\gamma_0(k)e^{ky+jkx} - \frac{jgA}{\omega}e^{ky-jkx}, & x \to +\infty, \tag{3.35}\\
\phi_0^{-\infty}(x, y) &\sim \left(\gamma_0(k) - \frac{jgA}{\omega}\right)e^{ky-jkx}, & x \to -\infty. \tag{3.36}
\end{align}
In (3.35), the leading-order spatial velocity potential as $x \to +\infty$ is the sum of the spatial velocity potential of the leading-order reflected wave and the spatial velocity potential of the incident wave:
\begin{equation}
\phi_0^{+\infty}(x, y) \sim \phi_{R0}(x, y) + \phi_I(x, y). \tag{3.37}
\end{equation}
From (3.37) and the linearized dynamic free surface boundary condition, the free surface elevation of leading-order reflected wave, $\eta_{R0}$, is calculated as below:
\begin{equation}
\eta_{R0} = -\frac{1}{g}\frac{\partial}{\partial t}(\Phi_{R0}(x, 0, t)) = -j\frac{\omega}{g}\gamma_0(k)e^{j(kx-\omega t)}. \tag{3.38}
\end{equation}
Since $\gamma_0(k)$ is predefined in (3.17) as,
\begin{equation}
\gamma_0(k) = C_0 a_1(k) - D_0 a_1''(k) = \frac{jgA}{\omega}\frac{\Delta_{11}}{\Delta_{123}}, \tag{3.39}
\end{equation}
the leading-order reflected wave elevation finally has the form of,
\begin{equation}
\eta_{R0} = -j\frac{\omega}{g}\frac{jgA}{\omega}\frac{\Delta_{11}}{\Delta_{123}}e^{j(kx-\omega t)} = A\frac{\Delta_{11}}{\Delta_{123}}e^{j(kx-\omega t)}. \tag{3.40}
\end{equation}
Thus, from (3.40), the leading-order reflected wave amplitude is found below:
\begin{equation}
A_{R0} = \left|A\frac{\Delta_{11}}{\Delta_{123}}\right| = A\frac{|\Delta_{11}|}{\sqrt{\Delta_{11}^2 + (\Delta_{12} - \Delta_{13})^2}}. \tag{3.41}
\end{equation}
Similarly, the leading-order spatial velocity potential as $x \to -\infty$ in (3.36) can be thought of as the spatial velocity potential of the leading-order transmitted wave:
\begin{equation}
\phi_0^{-\infty}(x, y) \sim \phi_{T0}(x, y). \tag{3.42}
\end{equation}
From (3.42) and the linearized dynamic free surface boundary condition, the leading-order transmitted wave is obtained as,
\begin{equation}
\eta_{T0} = -\frac{1}{g}\frac{\partial}{\partial t}\left(\Phi_{T0}^{\revB{-\infty}}(x, 0, t)\right) = A\frac{-j(\Delta_{12} - \Delta_{13})}{\Delta_{123}}e^{-j(kx+\omega t)}. \tag{3.43}
\end{equation}
Then, the wave amplitude of the leading-order transmitted wave will be,
\begin{equation}
A_{T0} = \left|A\frac{-j(\Delta_{12} - \Delta_{13})}{\Delta_{123}}\right| = A\frac{|\Delta_{12} - \Delta_{13}|}{\sqrt{\Delta_{11}^2 + (\Delta_{12} - \Delta_{13})^2}}. \tag{3.44}
\end{equation}
Dividing (3.41) and (3.44) \revB{by} the incident wave amplitude $A$, leading-order reflection coefficient and transmission coefficient can be obtained:
\begin{align}
R_0 &= \frac{A_{R0}}{A} = \frac{|\Delta_{11}|}{\sqrt{\Delta_{11}^2 + (\Delta_{12} - \Delta_{13})^2}}, \tag{3.45}\\
T_0 &= \frac{A_{T0}}{A} = \frac{|\Delta_{12} - \Delta_{13}|}{\sqrt{\Delta_{11}^2 + (\Delta_{12} - \Delta_{13})^2}}. \tag{3.46}
\end{align}
The reflection and transmission coefficients in (3.45) and (3.46) in various wave and breakwater conditions are plotted in figure 4. \revB{As a function of} $kb$ for different values of $a/b$, figure 4(a) illustrates the leading-order reflection coefficients, and figure 4(b) shows the leading-order transmission coefficients. \revB{Table 1 lists the computed values of $R_0$ and $T_0$ at representative points. Evans [8] gives the reflection and transmission coefficients of the impermeable plate in closed form (his equations (71)--(72)); on identifying $\Delta_{11} = a_1(-k)\,\gamma'$ and $\Delta_{12} - \Delta_{13} = a_1(-k)\,(\alpha' - \beta')$, where $\gamma'$, $\alpha'$, $\beta'$ are the integrals of [8, eq.~(69)], (3.45)--(3.46) are seen to be identical to Evans' expressions, and the two evaluations agree to machine precision at every point of table 1. The features of figure 2 of [8] are reproduced quantitatively: for $a/b = 0.25$ the maximum of $R_0$ is $0.175$ at $kb \approx 1.68$ (Evans: about $0.18$ at $kb = 1.7$), and for $a/b = 0.01$ the minimum of $T_0$ over $0.2 \le kb \le 4$ is $0.676$ at $kb \approx 2.7$ (Evans: $T$ never less than $0.68$). As the upper edge approaches the free surface, $R_0$ increases monotonically towards the surface-piercing value given in [8, \S7], but only logarithmically in $a/b$, as anticipated there: at $kb = 1$, $R_0 = 0.657$, $0.733$ and $0.811$ for $a/b = 10^{-3}$, $10^{-4}$ and $10^{-6}$, against the limiting value $0.947$.} \revB{The impermeable-plate values tabulated by Gayen \& Mondal [14, table 1] for a plate extending from $0.25p$ to $p$ below the surface, i.e.\ $a/b = 0.25$, differ from (3.45)--(3.46) by between $0.1\%$ and $4\%$ (table 2); the closed-form values coincide with Evans' expressions, with which [14] also reports graphical agreement.} \rev{Note also that (3.45)--(3.46) satisfy $R_0^2 + T_0^2 = 1$ identically, as they must for an impermeable plate.}

\begin{table}[htbp]
\centering
\revB{\caption{Leading-order (impermeable-plate) results: reflection and transmission coefficients (3.45)--(3.46), amplitude of the horizontal force (4.56) and centre of pressure (4.59). $R_0$, $T_0$ and $|\mathfrak{F}_0|$ coincide with Evans' equations (71)--(73) [8] to ten significant figures, and $c_{p0}$ with his equation (75).}
\begin{tabular}{cccccc}
\hline
$a/b$ & $kb$ & $R_0$ & $T_0$ & $|\mathfrak{F}_0|/(\rho g A b)$ & $c_{p0}$ \\
\hline
0.01 & 0.5 & 0.20728 & 0.97828 & 0.51700 & 0.4518 \\
0.01 & 1.0 & 0.52493 & 0.85115 & 0.78849 & 0.4315 \\
0.01 & 1.5 & 0.67431 & 0.73845 & 0.78792 & 0.4117 \\
0.01 & 2.0 & 0.72379 & 0.69002 & 0.72004 & 0.3927 \\
0.01 & 2.5 & 0.73652 & 0.67642 & 0.65032 & 0.3746 \\
0.05 & 0.5 & 0.15126 & 0.98849 & 0.38477 & 0.4687 \\
0.05 & 1.0 & 0.36958 & 0.92920 & 0.58014 & 0.4512 \\
0.05 & 1.5 & 0.48110 & 0.87667 & 0.60419 & 0.4341 \\
0.05 & 2.0 & 0.51701 & 0.85598 & 0.57001 & 0.4177 \\
0.05 & 2.5 & 0.51902 & 0.85476 & 0.52454 & 0.4021 \\
0.25 & 0.5 & 0.06639 & 0.99779 & 0.17972 & 0.4855 \\
0.25 & 1.0 & 0.14307 & 0.98971 & 0.25747 & 0.4732 \\
0.25 & 1.5 & 0.17314 & 0.98490 & 0.27151 & 0.4612 \\
0.25 & 2.0 & 0.17103 & 0.98527 & 0.25877 & 0.4495 \\
0.25 & 2.5 & 0.15427 & 0.98803 & 0.23673 & 0.4381 \\
\hline
\end{tabular}}
\end{table}

\begin{figure}[htbp]
\centering
\includegraphics[width=0.95\textwidth]{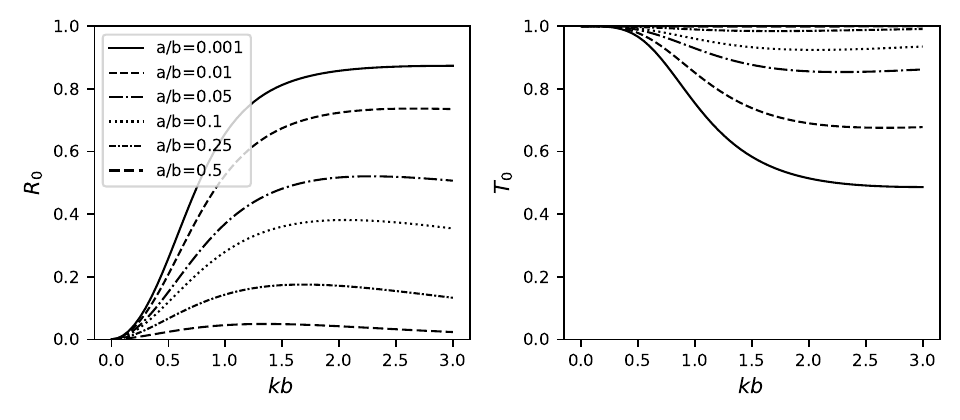}
\caption{Leading-order (a) reflection coefficient $R_0$ and (b) transmission coefficient $T_0$ versus $kb$ for $a/b = 0.001,\ 0.01,\ 0.05,\ 0.1,\ 0.25,\ 0.5$. \rev{(The authors' original figure (2023), recomputed with scripts written with Anthropic Claude (Fable~5); plotted values checked against figure 2 of [8].)}}
\end{figure}

\section{First-order solution}
\subsection{First-order velocity potential}
As proposed in \S2, it is assumed that the velocity potential comprises the perturbed sum of the leading-order and the first-order velocity potential, which represent the potential due to the impermeable plate and the permeable plate, respectively. In this section, the first-order correction for the case of a single permeable plate is sought.

In a similar way that the complex potential and the reduced potential are introduced to seek the solution for the leading-order problem, the first-order complex potential $w_1(z)$ and the first-order reduced potential $W_1(z)$ are proposed:
\begin{align}
w_1(z) &= \phi_1 + i\psi_1, & \Imi\{z\} < 0, \tag{4.1}\\
W_1(z) &= \frac{dw_1(z)}{dz} + ikw_1(z), & \Imi\{z\} < 0. \tag{4.2}
\end{align}
As in the previous section, the reduced potential $W_1(z)$ can be continued into $\Imi\{z\} > 0$ by Schwarz's reflection principle since the imaginary part of $W_1(z)$ is zero along the $x$-axis.

Firstly, the boundary condition on $L$ is sought before reflecting to $L'$. Using the notation of (4.2), the boundary condition of the first-order reduced potential on the permeable plate becomes,
\begin{equation}
\Rei\{W_1(z)\} = \frac{\partial\phi_1}{\partial x} - k\psi_1, \qquad \text{on } L. \tag{4.3}
\end{equation}
Unlike in the leading-order problem, the partial derivative of the spatial velocity potential is not zero (see (2.19)); thus, defining the boundary condition on the plate is necessary. In order to fully describe the boundary condition of the first-order reduced potential on the plate, the partial derivative of the first-order spatial potential $\phi_1$ with respect to $x$ and the first-order stream function must be derived. When the breakwater comprises a single permeable plate, the boundary condition of the first-order spatial velocity potential on the breakwater in (2.19) becomes,
\begin{equation}
\frac{\partial\phi_1}{\partial x} = -jk(\phi_0^+(0, y) - \phi_0^-(0, y)), \qquad \text{on } L. \tag{4.4}
\end{equation}
Taking the real part of the leading-order complex potential, which is presented in (3.33), the leading-order spatial velocity potential on the plate can be obtained:
\begin{equation}
\phi_0(x, y) = \Rei\left\{-\frac{jgA}{\omega}e^{-ikz}\left[1 - \frac{1}{\Delta_{123}}\int_{-ia}^{z} \frac{e^{ik\zeta}\left(a_1''(-k) + a_1(-k)\zeta^2\right)}{\sqrt{(\zeta^2 + a^2)(\zeta^2 + b^2)}}\, d\zeta\right]\right\}. \tag{4.5}
\end{equation}
Contracting the integration path onto the right and left sides of the plate, the leading-order spatial velocity potential on both sides of the plates can be calculated as follows:
\begin{equation}
\begin{split}
\phi_0^{\pm}(0, y) = -\frac{jgA}{\omega}e^{ky}\Bigg[1 \mp \frac{1}{\Delta_{123}}\Bigg(&\int_{-b}^{y} \frac{e^{-k\eta}\left(a_1''(-k) - a_1(-k)\eta^2\right)}{\sqrt{(\eta^2 - a^2)(b^2 - \eta^2)}}\, d\eta\\
&+ \int_{-a}^{-b} \frac{e^{-k\eta}\left(a_1''(-k) - a_1(-k)\eta^2\right)}{\sqrt{(\eta^2 - a^2)(b^2 - \eta^2)}}\, d\eta\Bigg)\Bigg].
\end{split}\tag{4.6}
\end{equation}
Defining
\begin{equation}
\xi_0(-k, y) = -\frac{jgA}{\omega}\frac{1}{\Delta_{123}}\int_{-b}^{y} \frac{e^{-k\eta}\left(a_1''(-k) - a_1(-k)\eta^2\right)}{\sqrt{(\eta^2 - a^2)(b^2 - \eta^2)}}\, d\eta, \tag{4.7}
\end{equation}
(4.6) is shortened to,
\begin{equation}
\phi_0^{\pm}(0, y) = e^{ky}\left[-\frac{jgA}{\omega} \mp \xi_0(-k, y)\right], \qquad \text{on } L. \tag{4.8}
\end{equation}
Substituting (4.8) into (4.4) gives a complete expression of the partial derivative of the first-order spatial velocity potential with respect to $x$:
\begin{equation}
\frac{\partial\phi_1}{\partial x} = j\,2ke^{ky}\xi_0(-k, y). \tag{4.9}
\end{equation}
To utilize (4.9) in finding the first-order stream function, the equation on the relation of the first-order spatial velocity potential and the first-order stream function \rev{is introduced}. Applying the Cauchy-Riemann equation \rev{to} (4.1),
\begin{equation}
\frac{\partial\phi_1}{\partial x} = \frac{\partial\psi_1}{\partial y}. \tag{4.10}
\end{equation}
\revB{When (4.10), with (4.9) substituted for its left-hand side, is integrated with respect to $y$ over $(-b, -a)$, $\psi_1(x, y)$ on the barrier is obtained.} Although previous studies [8, etc.] which dealt with impermeable breakwaters nicely defined $\psi_1(0, y)$ so that the constant of integration becomes zero, or just gave the indefinite integral function of $\frac{\partial\psi_1}{\partial y}$ with respect to $y$, the arbitrariness in $\phi_1$ should be resolved in this study since there is a flow through the permeable breakwater. Thus, this study defined the first-order stream function on $L$ as the sum of the definite integral of $\frac{\partial\psi_1}{\partial y}$ from $-b$ to $y$ and the constant of integration, $I$:
\begin{equation}
\psi_1(0, y) = \int_{-b}^{y} \left(j\,2ke^{k\eta}\xi_0(-k, \eta)\right) d\eta + I. \tag{4.11}
\end{equation}
Using integration by parts and substituting (4.7) into (4.11), $\psi_1(x, y)$ along the plate is obtained as a function of $y$:
\begin{equation}
\begin{split}
\psi_1(0, y) = j\,2e^{ky}\xi_0(-k, y) - \frac{gA}{\omega}\frac{2}{\Delta_{123}}\int_{-b}^{y} \frac{a_1''(-k) - a_1(-k)\eta^2}{\sqrt{(\eta^2 - a^2)(b^2 - \eta^2)}}\, d\eta + I,\\ \text{on } L.
\end{split} \tag{4.12}
\end{equation}
$\xi_0(-k, -b) = \gamma_0(-k) = 0$ on $L$, thus eliminated. Now, substituting (4.9) and (4.12) into (4.3), the boundary condition of the reduced potential at the permeable plate can be obtained:
\begin{equation}
\Rei\{W_1(z)\} = \frac{gA}{\omega}\frac{2k}{\Delta_{123}}\int_{-b}^{y} \frac{a_1''(-k) - a_1(-k)\eta^2}{\sqrt{(\eta^2 - a^2)(b^2 - \eta^2)}}\, d\eta - kI = f(y), \qquad \text{on } L. \tag{4.13}
\end{equation}
And this can be reflected in the upper-half complex plane:
\begin{equation}
\Rei\{W_1(z)\} = f(-|y|), \qquad \text{on } L + L'. \tag{4.14}
\end{equation}
Solving for $W_1(z)$ which satisfies (4.13) is a non-homogeneous Riemann-Hilbert problem, and the solution that vanishes at infinity can be found as,
\begin{equation}
W_1(z) = \frac{C_1 + D_1 z^2 + \frac{2}{\pi}\int_{-b}^{-a} \frac{\sqrt{(y^2 - a^2)(b^2 - y^2)}\,y f(y)}{y^2 + z^2}\, dy}{\sqrt{(z^2 + a^2)(z^2 + b^2)}}, \tag{4.15}
\end{equation}
where $C_1$, $D_1$ are some constants [26].

Similarly, from the leading order solution, the complex potential $w_1(z)$ can be obtained as,
\begin{equation}
w_1(z) = e^{-ikz}\left[B_1 + \int_{-ia}^{z} e^{ik\zeta}W_1(\zeta)\, d\zeta\right], \tag{4.16}
\end{equation}
where $B_1$ is a constant.

\rev{In particular, when} the integration path is taken to follow the plate, i.e., $y \in (-b, -a)$, the complex potential on the plate is as below:
\begin{equation}
w_1^{\pm}(iy) = e^{ky}\left[B_1 - \int_{-a}^{y} e^{-ku}\Imi\left\{W_1^{\pm}(iu)\right\} du + i\int_{-a}^{y} e^{-ku}f(u)\, du\right]. \tag{4.17}
\end{equation}
Taking the imaginary part of (4.17) only, the first-order stream function on the plate is obtained:
\begin{equation}
\psi_1(0, y) = \Imi\left\{w_1^{\pm}(iy)\right\} = e^{ky}\int_{-a}^{y} e^{-ku}f(u)\, du, \qquad \text{on } L. \tag{4.18}
\end{equation}
Here, $\psi_1(0, y)$ in (4.12) and $\psi_1(0, y)$ in (4.18) should be the same. In other words, the integral of $\phi(0, y)$ along the plate and the imaginary part of $w_1(z)$ on the plate must have the same value. Hence, one equation for determining the constant of integration, $I$, is presented as:
\begin{equation}
\begin{split}
&j\,2e^{ky}\xi_0(-k, y) - \frac{gA}{\omega}\frac{2}{\Delta_{123}}\int_{-b}^{y} \frac{a_1''(-k) - a_1(-k)\eta^2}{\sqrt{(\eta^2 - a^2)(b^2 - \eta^2)}}\, d\eta + I\\
&\qquad = e^{ky}\int_{-a}^{y} e^{-ku}\left(\frac{gA}{\omega}\frac{2k}{\Delta_{123}}\int_{-b}^{u} \frac{a_1''(-k) - a_1(-k)\eta^2}{\sqrt{(\eta^2 - a^2)(b^2 - \eta^2)}}\, d\eta - kI\right) du.
\end{split} \tag{4.19}
\end{equation}
When $y \to -b$,
\begin{equation}
I = \frac{gA}{\omega}\frac{2k}{\Delta_{123}}e^{-kb}\int_{-a}^{-b} e^{-ku}\int_{-b}^{u} \frac{a_1''(-k) - a_1(-k)\eta^2}{\sqrt{(\eta^2 - a^2)(b^2 - \eta^2)}}\, d\eta\, du + \left(1 - e^{k(a-b)}\right)I. \tag{4.20}
\end{equation}
Therefore, rearranging the equation above gives the value of $I$:
\begin{equation}
I = \frac{gA}{\omega}\frac{2k}{\Delta_{123}}e^{-ka}\int_{-a}^{-b} e^{-ku}\int_{-b}^{u} \frac{a_1''(-k) - a_1(-k)\eta^2}{\sqrt{(\eta^2 - a^2)(b^2 - \eta^2)}}\, d\eta\, du. \tag{4.21}
\end{equation}
Thus, $f(y)$ is now fully defined by substituting the expression of $I$ in (4.21) into (4.13):
\begin{equation}
f(y) = \frac{gA}{\omega}\frac{2k}{\Delta_{123}}\left[\int_{-b}^{y} \frac{a_1''(-k) - a_1(-k)\eta^2}{\sqrt{(\eta^2 - a^2)(b^2 - \eta^2)}}\, d\eta - k\tilde{I}\right] = \frac{gA}{\omega}\tilde{f}(y). \tag{4.22}
\end{equation}
In (4.22), it is worth noting that the first integral term can be expressed in elliptic integral form.
\begin{equation}
\int_{-y}^{b} \frac{a_1''(-k) - a_1(-k)\eta^2}{\sqrt{(\eta^2 - a^2)(b^2 - \eta^2)}}\, d\eta = \frac{a_1''(-k)}{b}F(\varphi, m) - a_1(-k)\,b\,E(\varphi, m), \tag{4.23}
\end{equation}
where $F(\varphi, m)$ is the incomplete elliptic integral of the first kind and $E(\varphi, m)$ is the incomplete elliptic integral of the second kind [30, p. 56], and each parameter represents,
\begin{equation}
\varphi = \sin^{-1}\sqrt{\frac{b^2 - y^2}{b^2 - a^2}}, \qquad m = \frac{b^2 - a^2}{b^2}. \tag{4.24}
\end{equation}
Using these continuous elliptic integral functions and the previously calculated value of $I$, $f(y)$ with various $kb$ and $a/b$ can be plotted as shown in figure 5.

\begin{figure}[htbp]
\centering
\includegraphics[width=0.95\textwidth]{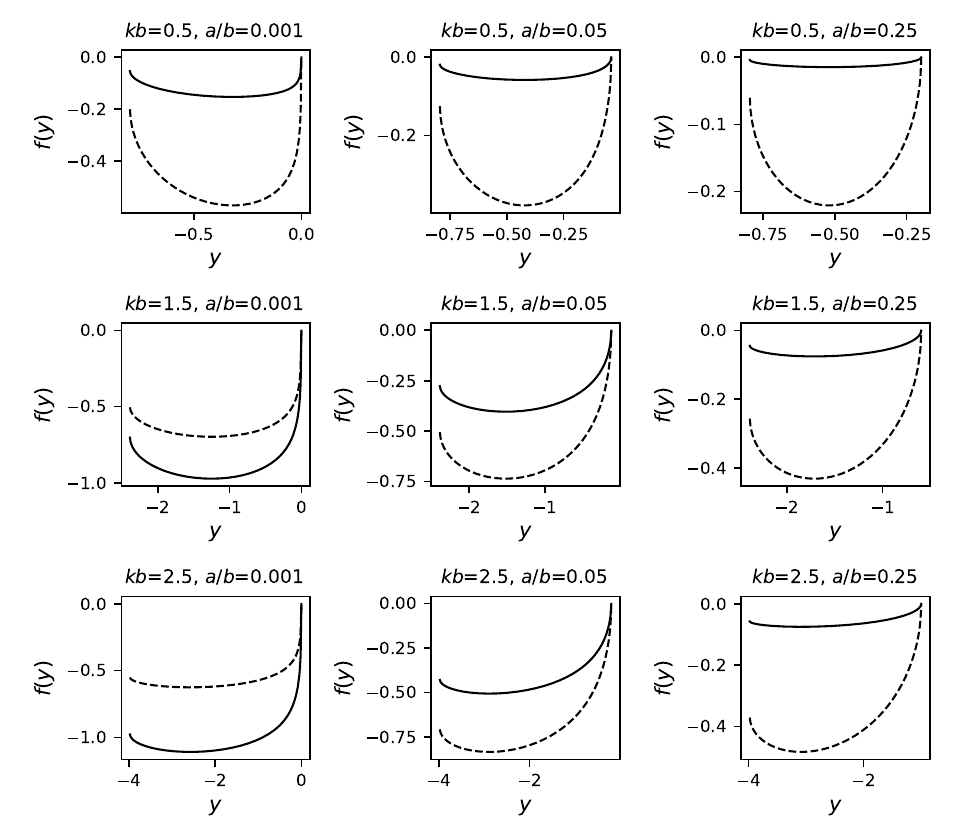}
\caption{$f(y)$ versus $y$ for (top) $kb = 0.5$, (center) $kb = 1.5$, (bottom) $kb = 2.5$; (a) $a/b = 0.001$, (b) $a/b = 0.05$, (c) $a/b = 0.25$. (---, $\Rej\{f(y)\}$; - - -, $\Imj\{f(y)\}$.) \rev{(The authors' original figure (2023), recomputed with scripts written with Anthropic Claude (Fable~5); results verified via the energy identity (5.4).)}}
\end{figure}

To define the unknown constants $B_1$, $C_1$, and $D_1$, the same way used in the leading-order problem can be utilized. First, the \revB{radiation} boundary conditions will be applied. Since $W_1(z) - D_1 \to 0$ as $|z| \to \pm\infty$, for large $z$,
\begin{equation}
w_1(z) \sim e^{-ikz}\left[B_1 + \int_{-ia}^{i\infty} e^{ik\zeta}W_1(\zeta)\, d\zeta - \frac{iD_1}{k}e^{ikz}\right]. \tag{4.25}
\end{equation}
As in the previous section, the integration path along the $y$-axis is split into three intervals by defining functions as below:
\begin{align}
F(u) &= \frac{gA}{\omega}\frac{2}{\pi}\int_{-b}^{-a} \frac{\sqrt{(y^2 - a^2)(b^2 - y^2)}\,y\tilde{f}(y)}{y^2 - u^2}\, dy = \frac{gA}{\omega}\tilde{F}(u), \tag{4.26}\\
a_1(k, F) &= \frac{gA}{\omega}\int_a^b \frac{e^{-ku}\tilde{F}(u)}{\sqrt{(u^2 - a^2)(b^2 - u^2)}}\, du = \frac{gA}{\omega}a_1(k, \tilde{F}), \tag{4.27}\\
a_2(k, F) &= \frac{gA}{\omega}\int_{-a}^{a} \frac{e^{-ku}\tilde{F}(u)}{\sqrt{(a^2 - u^2)(b^2 - u^2)}}\, du = \frac{gA}{\omega}a_2(k, \tilde{F}), \tag{4.28}\\
a_3(k, F) &= \frac{gA}{\omega}\int_b^{\infty} \frac{e^{-ku}\tilde{F}(u)}{\sqrt{(u^2 - a^2)(u^2 - b^2)}}\, du = \frac{gA}{\omega}a_3(k, \tilde{F}), \tag{4.29}\\
\gamma_1(k) &= C_1 a_1(k) - D_1 a_1''(k) + a_1(k, F), \tag{4.30}\\
\alpha_1(k) &= C_1 a_2(k) - D_1 a_2''(k) + a_2(k, F), \tag{4.31}\\
\beta_1(k) &= C_1 a_3(k) - D_1 a_3''(k) + a_3(k, F), \tag{4.32}\\
\delta_1(k) &= \frac{gA}{\omega}\int_a^b e^{-ku}\tilde{f}(u)\, du = \frac{gA}{\omega}\tilde{\delta}_1(k). \tag{4.33}
\end{align}
Then, (4.16) can be simplified:
\begin{equation}
w_1(z) \sim e^{-ikz}\left[B_1 \mp \gamma_1(k) + i\left(\alpha_1(k) - \beta_1(k) + \delta_1(k)\right)\right] - \frac{iD_1}{k}. \tag{4.34}
\end{equation}
Around the plate, the zero-circulation condition is enforced:
\begin{equation}
\Rei\left\{\oint_{\Gamma} e^{iku}W_1(u)\, du\right\} = 0. \tag{4.35}
\end{equation}
(4.35) can be shortened using the notations above:
\begin{equation}
\gamma_1(-k) = C_1 a_1(-k) - D_1 a_1''(-k) + a_1(-k, F) = 0. \tag{4.36}
\end{equation}
We also defined the radiation boundary conditions at $x \to \pm\infty$ in (2.22) and (2.23). Substituting the first-order complex potential (4.34) into these radiation boundary conditions gives,
\begin{equation}
\gamma_1(k) = j\left(\alpha_1(k) - \beta_1(k) + \delta_1(k)\right). \tag{4.37}
\end{equation}
Combining (4.36) and (4.37), $B_1$, $C_1$, and $D_1$ are obtained to be,
\begin{align}
B_1 &= 0, \tag{4.38}\\
C_1 &= \frac{\Delta_{F1}'' - j\left(\Delta_{F2}'' - \Delta_{F3}'' - a_1''(-k)\delta_1(k)\right)}{\Delta_{123}}, \tag{4.39}\\
D_1 &= \frac{\Delta_{F1} - j\left(\Delta_{F2} - \Delta_{F3} - a_1(-k)\delta_1(k)\right)}{\Delta_{123}}, \tag{4.40}
\end{align}
where,
\begin{align}
\Delta_{Fi} &= \begin{vmatrix} a_i(k) & a_1(-k)\\ a_i(k, F) & a_1(-k, F) \end{vmatrix}, \qquad i = 1, 2, 3, \tag{4.41}\\
\Delta_{Fi}'' &= \begin{vmatrix} a_i''(k) & a_1''(-k)\\ a_i(k, F) & a_1(-k, F) \end{vmatrix}, \qquad i = 1, 2, 3. \tag{4.42}
\end{align}
Now, the first-order solution of the complex potential is fully defined, and the spatial potential is also obtained \revB{by} taking the real part of the complex potential:
\begin{equation}
\phi_1(x, y) = \Rei\{w_1(z)\}. \tag{4.43}
\end{equation}

\subsection{Wave scattering by a permeable plate}
In order to consider the effect of the permeability of the plate, \rev{the first-order corrections for a reflection coefficient and a transmission coefficient} are presented in this section.

Using the constants we obtained, the first-order spatial wave potential at $x \to +\infty$ is calculated as below:
\begin{equation}
\phi_1^{+\infty}(x, y) \sim -\gamma_1(k)e^{jkx+ky} \sim \phi_{R1}^{+\infty}(x, y). \tag{4.44}
\end{equation}
To distinguish the real and imaginary \revB{parts} of $\gamma_1(k)$ with respect to $j$, we rearranged the expression of $\gamma_1(k)$:
\begin{equation}
\gamma_1(k) = C_1 a_1(k) - D_1 a_1''(k) + a_1(k, F) = -\frac{jgA}{\omega}\frac{\Lambda}{\Delta_{123}}, \tag{4.45}
\end{equation}
where,
\begin{align}
\Delta_i &= \begin{vmatrix} a_i(k) & a_1(k)\\ a_i''(k) & a_1''(k) \end{vmatrix}, \qquad i = 2, 3, \tag{4.46}\\
\begin{split}
\Lambda &= a_1(k, \tilde{F})(\Delta_{12} - \Delta_{13}) - a_1(-k, \tilde{F})(\Delta_2 - \Delta_3)\\
&\quad - (a_2(k, \tilde{F}) - a_3(k, \tilde{F}) + \tilde{\delta}_1(k))\Delta_{11}.
\end{split} \tag{4.47}
\end{align}
With this simplified expression, the first-order reflected wave at $x \to \infty$ \rev{is obtained} below:
\begin{equation}
\eta_{R1} = -\frac{1}{g}\frac{\partial}{\partial t}\left(\Phi_{R1}^{+\infty}(x, 0, t)\right) = -A\frac{\Lambda}{\Delta_{123}}e^{j(kx-\omega t)}. \tag{4.48}
\end{equation}
Likewise, the first-order spatial velocity potential at $x \to -\infty$ is,
\begin{equation}
\phi_1^{-\infty}(x, y) \sim \gamma_1(k)e^{-jkx+ky} \sim \phi_{T1}^{-\infty}(x, y). \tag{4.49}
\end{equation}
From the result above, the first-order transmitted wave can be obtained.
\begin{equation}
\eta_{T1} = -\frac{1}{g}\frac{\partial}{\partial t}\left(\Phi_{T1}^{\revB{-\infty}}(x, 0, t)\right) = A\frac{\Lambda}{\Delta_{123}}e^{-j(kx+\omega t)}. \tag{4.50}
\end{equation}
Consequently, from the perturbed solution form as in (2.17), the reflected wave and the transmitted wave are expressed in the perturbed series of the form,
\begin{align}
\eta_R &= \eta_{R0} + \varepsilon\eta_{R1} = A\frac{(\Delta_{11} - \varepsilon\Lambda)}{\Delta_{123}}e^{j(kx-\omega t)}, \tag{4.51}\\
\eta_T &= \eta_{T0} + \varepsilon\eta_{T1} = -A\frac{(j(\Delta_{12} - \Delta_{13}) - \varepsilon\Lambda)}{\Delta_{123}}e^{-j(kx+\omega t)}. \tag{4.52}
\end{align}
Dividing (4.51) and (4.52) \revB{by} the incident wave amplitude $A$, the reflection coefficient and transmission coefficient considering the permeability of the plate are represented below.
\begin{align}
R &= \frac{A_R}{A} = \frac{|\Delta_{11} - \varepsilon\Lambda|}{\sqrt{\Delta_{11}^2 + (\Delta_{12} - \Delta_{13})^2}}, \tag{4.53}\\
T &= \frac{A_T}{A} = \frac{|j(\Delta_{12} - \Delta_{13}) - \varepsilon\Lambda|}{\sqrt{\Delta_{11}^2 + (\Delta_{12} - \Delta_{13})^2}}. \tag{4.54}
\end{align}

\revB{\subsection{Potential jump on the plate and hydrodynamic load}
The load on the breakwater, unlike the far-field amplitudes of \S4.2, requires the first-order potential on the plate itself. On the cut $z = iu$, $u \in (-b, -a)$, the square root in (4.15) takes the values $\pm i\sqrt{(u^2 - a^2)(b^2 - u^2)}$ on the two faces, with the same convention as in (4.6), so that (4.17) with $B_1 = 0$ gives the jump of the first-order potential across the plate, for $y$ on $L$,
\begin{equation}
\phi_1^+(0, y) - \phi_1^-(0, y) = -2e^{ky}\int_{y}^{-a} \frac{e^{-ku}\left(C_1 - D_1 u^2 + F(-u)\right)}{\sqrt{(u^2 - a^2)(b^2 - u^2)}}\, du. \tag{4.55}
\end{equation}
The jump vanishes at the upper edge identically and at the lower edge by (4.36), as it must for a potential that is continuous around the edges; the leading-order jump is $\phi_0^+ - \phi_0^- = -2e^{ky}\xi_0(-k, y)$ by (4.8), and the same expression is recovered from (4.55) on replacing $(C_1, D_1, F)$ by $(C_0, D_0, 0)$.

With the linearized pressure (2.13), the hydrostatic part cancels across the plate and the horizontal force per unit width on the plate, positive in the $+x$ direction, is
\begin{equation}
\begin{split}
F_x(t) &= \int_L (p^- - p^+)\, dy = \Rej\left\{-j\omega\rho\, \mathfrak{F}\, e^{-j\omega t}\right\},\\
\mathfrak{F} &= \int_{-b}^{-a} (\phi^+ - \phi^-)\, dy = \mathfrak{F}_0 + \varepsilon\,\mathfrak{F}_1,
\end{split} \tag{4.56}
\end{equation}
with
\begin{equation}
\begin{split}
\mathfrak{F}_0 &= -2\int_{-b}^{-a} e^{ky}\xi_0(-k, y)\, dy,\\
\mathfrak{F}_1 &= -\frac{2}{k}\int_{-b}^{-a} \frac{e^{-ku}\left(C_1 - D_1 u^2 + F(-u)\right)}{\sqrt{(u^2 - a^2)(b^2 - u^2)}}\left(e^{ku} - e^{-kb}\right) du,
\end{split} \tag{4.57}
\end{equation}
where the order of integration in (4.55) has been exchanged. The moment about the origin, positive in the sense that carries the $+x$ direction towards $+y$, has the amplitude
\begin{equation}
\mathfrak{M} = \int_{-b}^{-a} y\,(\phi^+ - \phi^-)\, dy = \mathfrak{M}_0 + \varepsilon\,\mathfrak{M}_1, \tag{4.58}
\end{equation}
with $\mathfrak{M}_0$, $\mathfrak{M}_1$ obtained from (4.57) on inserting the factor $y$ before the inner integration, and the centre of pressure, measured downward from the upper edge as a fraction of the plate length, is
\begin{equation}
c_p = \frac{-a - \Rej\{\mathfrak{M}/\mathfrak{F}\}}{b - a}. \tag{4.59}
\end{equation}
The integrals in (4.57) have the same inverse-square-root endpoint behaviour as (4.27) and are evaluated by the same Gauss--Chebyshev rule, reusing the values of $F$ at its nodes.

In the impermeable limit $\varepsilon = 0$, $\mathfrak{F}_0$ and $c_{p0}$ must reduce to the first-order force and centre of pressure obtained by Evans [8, eqs.~(73) and (75)] in terms of the complete elliptic integrals $K(\kappa)$, $E(\kappa)$ of modulus $\kappa = (1 - a^2/b^2)^{1/2}$. The direct quadrature of (4.57) and Evans' elliptic form agree to ten significant figures at every point of table 1; since the elliptic reduction enters Evans' result but not (4.57), this is an independent check of the leading-order potential jump (4.8), which is the forcing of the whole first-order problem. At first order, the change of the force amplitude is $d|\mathfrak{F}|/d\varepsilon|_{\varepsilon=0} = 2\Rej\{\bar{\mathfrak{F}}_0\mathfrak{F}_1\}/|\mathfrak{F}_0|$; over the whole grid of \S6 ($0.05 \le kb \le 2.5$, $a/b = 0.001$, $0.05$, $0.25$) the phase of $\mathfrak{F}_1/\mathfrak{F}_0$ lies between $-90^\circ$ and $-148^\circ$, so that this derivative is negative and the permeability relieves the load at first order, as in the porous-wavemaker theory [11]. The correction is, however, nearly in quadrature with the leading-order force. This is to be expected, since by (2.15) the seepage velocity that forces the first-order problem is itself in quadrature with the leading-order pressure jump. The term quadratic in $\varepsilon$ in $|\mathfrak{F}_0 + \varepsilon\mathfrak{F}_1|^2$, which is not compensated at this order of truncation, therefore becomes important at moderate $\varepsilon$. The truncated amplitude falls below the impermeable value only for $\varepsilon < \varepsilon_F = -2\Rej\{\bar{\mathfrak{F}}_0\mathfrak{F}_1\}/|\mathfrak{F}_1|^2$, and is quantitatively meaningful only for $\varepsilon$ well below $\varepsilon_F$; this bound is considerably stricter than (5.6) (see \S6). The closure of (4.55) at the lower edge, $\int_{-b}^{-a} e^{-ku}(C_1 - D_1u^2 + F(-u))/\sqrt{(u^2-a^2)(b^2-u^2)}\, du = 0$, is satisfied to a relative residual below $10^{-10}$ at every point of the grid, which checks the constants (4.39)--(4.40).}

\rev{\section{Energy identity and \revB{the energy-consistency range} of the perturbation expansion}
\subsection{An exact energy identity}
The porous boundary condition (2.15) is dissipative, and energy conservation provides an exact constraint on the scattering coefficients that the solution must satisfy. When Green's identity is applied to $\phi$ and its complex conjugate $\bar{\phi}$ over the fluid domain bounded by the free surface, two vertical control lines at $x \to \pm\infty$, and the two faces of the plate, the free-surface contribution vanishes by (2.3), and the far-field contributions reduce to the wave-energy fluxes. On the plate faces the normal velocity is continuous, and eliminating it with (2.15) yields a strictly non-negative contribution proportional to $|\phi^+ - \phi^-|^2$. The result is the exact identity
\begin{equation}
R^2 + T^2 = 1 - \frac{2k\varepsilon}{(gA/\omega)^2}\int_L |\phi^+ - \phi^-|^2\, dy, \tag{5.1}
\end{equation}
valid for the full solution of (2.2)--(2.15) at any $\varepsilon \ge 0$: the wave energy not reflected or transmitted is dissipated by the pressure-driven flow through the plate.

\subsection{The identity at first order}
Writing the complex reflection and transmission amplitudes from (4.51)--(4.52) as $r = r_0 + \varepsilon r_1$ and $t = t_0 + \varepsilon t_1$ with
\begin{equation}
r_0 = \frac{\Delta_{11}}{\Delta_{123}}, \quad t_0 = \frac{-j(\Delta_{12} - \Delta_{13})}{\Delta_{123}}, \quad r_1 = -\frac{\Lambda}{\Delta_{123}}, \quad t_1 = \frac{\Lambda}{\Delta_{123}}, \tag{5.2}
\end{equation}
and noting $|r_0|^2 + |t_0|^2 = 1$, one finds that the $O(\varepsilon)$ balance of (5.1) requires
\begin{equation}
2\,\Rej\{\bar{r}_0 r_1 + \bar{t}_0 t_1\} = -\mathcal{D}_0, \qquad \mathcal{D}_0 = \frac{8k}{(gA/\omega)^2}\int_{-b}^{-a} e^{2ky}\,|\xi_0(-k, y)|^2\, dy, \tag{5.3}
\end{equation}
where the leading-order potential jump across the plate is $\phi_0^+ - \phi_0^- = -2e^{ky}\xi_0(-k, y)$ by (4.8). Substituting (5.2) into (5.3) and using $\Delta_{123} = \Delta_{11} - j(\Delta_{12} - \Delta_{13})$ gives the equivalent statement
\begin{equation}
\frac{\Rej\{\Lambda\,\Delta_{123}\}}{|\Delta_{123}|^2} = \frac{\mathcal{D}_0}{2} \ \ge 0. \tag{5.4}
\end{equation}
Equation (5.4) is a nontrivial identity connecting the first-order far-field quantity $\Lambda$ of (4.47) to a weighted norm of the leading-order solution on the plate. It has been verified numerically to hold to within the quadrature tolerance (relative error below \revB{$10^{-7}$}) over the whole range $0.3 \le kb \le 2.5$ and the values of $a/b$ presented in this paper, which provides an independent check of the first-order analysis of \S4.

\subsection{Closed-form \revB{energy-consistency bound on} the expansion}
For the truncated coefficients (4.53)--(4.54) an elementary computation gives, exactly,
\begin{equation}
R^2 + T^2 = 1 - 2\varepsilon\,\frac{\Rej\{\Lambda\Delta_{123}\}}{|\Delta_{123}|^2} + 2\varepsilon^2\frac{|\Lambda|^2}{|\Delta_{123}|^2} = 1 - \varepsilon\,\mathcal{D}_0 + 2\varepsilon^2\frac{|\Lambda|^2}{|\Delta_{123}|^2}, \tag{5.5}
\end{equation}
where the second equality uses (5.4). The $O(\varepsilon)$ term is the physical dissipation; the positive $O(\varepsilon^2)$ term is an artefact of the truncation, uncompensated because the second-order field is not included. The truncated solution therefore remains energy-consistent, $R^2 + T^2 \le 1$, if and only if
\begin{equation}
\varepsilon \ \le\ \varepsilon_{\max}(kb;\ a/b) = \frac{\Rej\{\Lambda\,\Delta_{123}\}}{|\Lambda|^2}. \tag{5.6}
\end{equation}
This closed-form \revB{bound} replaces any ad hoc restriction of the plotted range: since $|\Lambda|$ grows with $kb$ while the numerator remains bounded, $\varepsilon_{\max}$ decreases monotonically with $kb$ (figure 6), and the apparent divergence of $R$ and $T$ at large $kb$ reported for fixed $\varepsilon$ is simply the \revB{truncated} expansion \revB{ceasing to be energy-consistent}. At the boundary $\varepsilon = \varepsilon_{\max}$ the magnitude of the first-order correction to the far-field amplitudes reaches the order of the incident wave itself, which is the classical signature of the breakdown of a regular perturbation expansion. \revB{Equation (5.6) is a necessary condition for the truncated expansion to be physically meaningful, obtained from energy consistency alone; it is not a proof of the convergence of the expansion, and an error estimate for the truncation would require the second-order problem, which is not attempted here. For this reason $\varepsilon_{\max}$ is referred to throughout as the energy-consistency bound of the expansion.}

Physically, $\varepsilon \le \varepsilon_{\max}$ delimits the regime in which the wave perceives the plate as a nearly solid barrier with a weak seepage through-flow. In engineering variables, $\varepsilon = \kappa\omega/(D\nu k)$ decreases for finer pores (smaller $\kappa$) and thicker plates (larger $D$). For a fixed plate in deep water, $\varepsilon = \kappa\sqrt{g}/(D\nu\sqrt{k})$ grows slowly, like $k^{-1/2}$, as the incident wave becomes longer, whereas figure 6 shows that $\varepsilon_{\max}$ grows much faster in the long-wave limit $kb \to 0$ (approximately like $(kb)^{-3}$ for the geometries considered); the criterion (5.6) is therefore increasingly well satisfied for long waves, and it is at the short-wave end, where $\varepsilon_{\max}$ collapses while $\varepsilon$ decreases only like $k^{-1/2}$, that the \revB{truncated} expansion \revB{ceases to be energy-consistent}. For a fixed plate the \revB{energy-consistency range} in wave-period space is thus one-sided, and (5.6) provides the quantitative criterion to be checked in applications.}

\revB{The porosity parameter has been taken real, corresponding to a purely resistive through-flow governed by (2.11). Inertial effects in the pores are commonly represented by a complex porous-effect parameter [11, 14]. Formally, the analysis of \S\S3--4 goes through unchanged for complex $\varepsilon$, since $\varepsilon$ enters only as a multiplier in (2.19) and the first-order far-field amplitudes remain $\varepsilon r_1$ and $\varepsilon t_1$ with $r_1$ and $t_1$ as in (5.2); the dissipative term in (5.1), however, then involves the real part of $\varepsilon$ only, and the bound (5.6) must be re-derived accordingly. A systematic treatment of this case, and of porosity parameters that are not small, lies outside the one-sided expansion about the impermeable plate and is left for future work.}

\begin{figure}[htbp]
\centering
\includegraphics[width=0.6\textwidth]{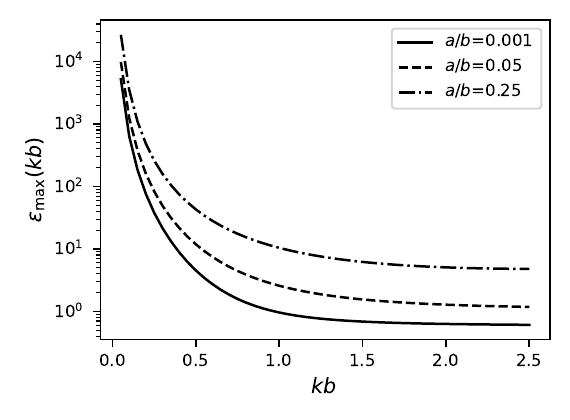}
\caption{\rev{\revB{Energy-consistency bound} $\varepsilon_{\max}(kb) = \Rej\{\Lambda\Delta_{123}\}/|\Lambda|^2$ of the truncated perturbation expansion for $a/b = 0.001,\ 0.05,\ 0.25$. (Script written with Anthropic Claude (Fable~5) under the authors' direction; results verified via the energy identity (5.4).)}}
\end{figure}

\section{Numerical evaluation and results}
\subsection{Numerical approximate integration of $R$ and $T$}
Unlike in the case of the leading-order problem, the reflection and transmission coefficients of the first-order solution for the velocity potential cannot be easily evaluated in the ordinary sense \revB{since $F(u)$ itself involves an improper integral.} In (4.26), in the denominator of the integrand, $u^2$ can have the same value \revB{as} $y^2$ when $y \in (-b, -a)$ and $u \in (a, b)$. Thus, this integral cannot be calculated in the normal quadrature rule, and it has the same form of finite Hilbert transform---or Cauchy principal value integral---of $\sqrt{(y^2 - a^2)(b^2 - y^2)}\,yf(y)$. This type of singular integral equation can be seen in numerous practical problems in science and engineering disciplines [31, 32]. Therefore, an appropriate scheme for the numerical evaluation of this integral is needed.

In this paper, following the uniform approximation methods to finite Hilbert transform by [33], $F(u)$ \revB{was calculated numerically by interpolating the numerator with Chebyshev polynomials}. First of all, (4.26) is transformed \revB{by} \emph{subtracting out the singularity}, which is the typical procedure for approximating the principal value integral [31, 33--37]. Let the Cauchy principal value of $F(u)$ \rev{be} $\frac{1}{\pi}I(g; u^2)$, where $g$ represents the numerator of the integrand in $F(u)$.
\begin{equation}
\begin{split}
I(g; u^2) &= 2\,\mathrm{p.v.}\int_{-b}^{-a} \frac{\sqrt{(y^2 - a^2)(b^2 - y^2)}\,yf(y)}{y^2 - u^2}\, dy\\
&= \int_{b^2}^{a^2} \frac{g(\tau) - g(u^2)}{\tau - u^2}\, d\tau + g(u^2)\ln\left(\frac{a^2 - u^2}{u^2 - b^2}\right).
\end{split} \tag{6.1}
\end{equation}
(6.1) may be evaluated by approximating $g(\tau)$ as a polynomial $p_M(t)$, which is the finite sum of Chebyshev \revB{polynomials} of the first kind $T_r(t)$ [36]. Here, $\tau$ is the variable \rev{such that} the interval of $t$, $[-1, 1]$, is mapped to $[b^2, a^2]$ by the affine transformation.
\begin{equation}
\tau = \frac{a^2 - b^2}{2}t + \frac{a^2 + b^2}{2}, \qquad t \in [-1, 1]. \tag{6.2}
\end{equation}
When $t$ is given as $t = \cos\theta$, $T_r(t) = \cos r\theta$. Then, the function $g(\tau)$ can be approximated \revB{by} $p_M(t)$ as below [38].
\begin{equation}
p_M(t) = \sum_{r=0}^{M}{}^{\prime\prime} a_r^M T_r(t), \qquad -1 \le t \le 1, \tag{6.3}
\end{equation}
where double prime is the symbol of the summation whose first and last terms are halved. Selecting $t_q = \cos(\pi q/M)$, $q = 0, \cdots, M$ as the $M + 1$ interpolating \revB{abscissae}, the interpolation condition becomes,
\begin{equation}
g(\tau_q) = p_M(t_q) = p_M\left(\cos\left(\frac{\pi q}{M}\right)\right), \qquad q = 0, \cdots, M. \tag{6.4}
\end{equation}
Following the quadrature method of Hasegawa [39], (6.1) is evaluated by substituting $p_M(t)$ into $g(\tau)$ and adopting the quadrature rule using Chebyshev expansion in terms of $T_r(t)$. \revB{Once $F(u)$ has been found from (4.26), $a_1(k, F)$ and $\gamma_1(k)$ follow from (4.27) and (4.30), and the reflection and transmission coefficients of the permeable plate can be evaluated.}

\rev{The numerical evaluation was implemented in Python scripts written with Anthropic Claude (Fable~5) under the authors' direction, and the first-order solution computed in this way satisfies the exact energy identity derived in \S5 to within the quadrature tolerance over the whole range of $kb$ presented; the scripts are archived with the figure data [40].}

Before computing the reflection and transmission coefficients, the convergence of $a_M^M$ was examined. $a_M^M$, the last coefficient of the approximating polynomial weighs heavily when calculating the upper limit of error. \revB{Thus $M$ was chosen so that $|a_M^M|$ fell below the tolerance $10^{-6}$.} Therefore, using $M + 1 = 2^{12} + 1$ points, $F(u)$ was evaluated. \revB{The profile of $F(u)$ is shown in figure 7.}

\begin{figure}[htbp]
\centering
\includegraphics[width=0.95\textwidth]{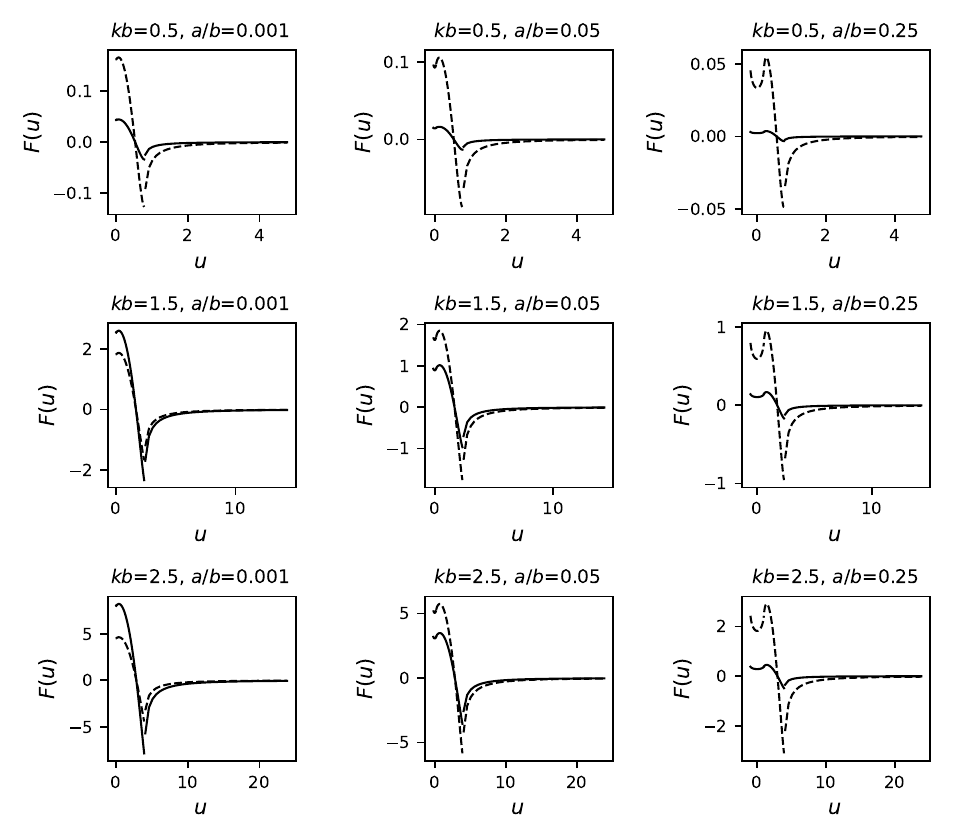}
\caption{$F(u)$ versus $u$ for (top) $kb = 0.5$, (center) $kb = 1.5$, (bottom) $kb = 2.5$; (a) $a/b = 0.001$, (b) $a/b = 0.05$, (c) $a/b = 0.25$. (---, $\Rej\{F(u)\}$; - - -, $\Imj\{F(u)\}$.) \rev{(The authors' original figure (2023), recomputed with scripts written with Anthropic Claude (Fable~5); results verified via the energy identity (5.4).)}}
\end{figure}

$\Lambda$ in (4.47) determines the magnitude of the first-order term in reflection and transmission coefficients, and critically affects the applicable range of the perturbation method. \rev{Figure 8 shows the recomputed $\Lambda$: its magnitude grows with $kb$, so that for a fixed $\varepsilon$ the first-order correction eventually ceases to be small compared with the leading order. Instead of restricting the plotted range in an ad hoc manner, the range in which the truncated expansion remains energy-consistent is quantified by the closed-form \revB{bound} $\varepsilon_{\max}(kb)$ derived in \S5; in \revB{figure} 9 the curves are drawn in grey beyond this boundary, and the vertical lines mark the value $kb^*$ at which $\varepsilon = \varepsilon_{\max}(kb^*)$ for each $\varepsilon$.}

\begin{figure}[htbp]
\centering
\includegraphics[width=0.95\textwidth]{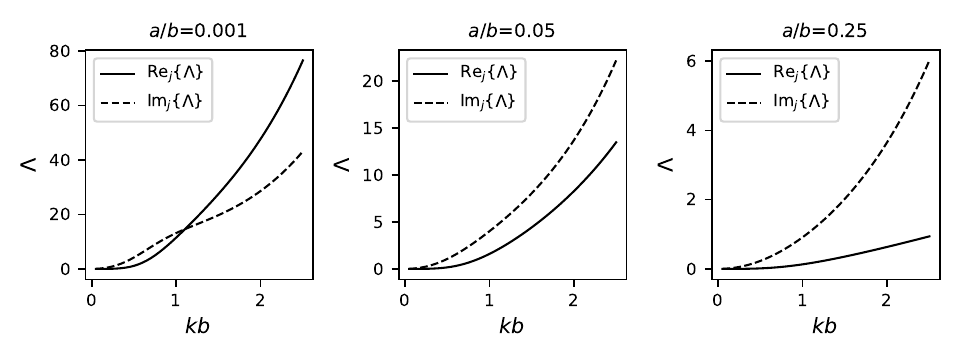}
\caption{$\Lambda$ versus $kb$ for (a) $a/b = 0.001$, (b) $a/b = 0.05$, (c) $a/b = 0.25$. (---, $\Rej\{\Lambda\}$; - - -, $\Imj\{\Lambda\}$.) \rev{(The authors' original figure (2023), recomputed with scripts written with Anthropic Claude (Fable~5); results verified via the energy identity (5.4).)}}
\end{figure}

Finally, the reflection and transmission coefficients considering the permeable effect are presented. In figure 9, it is seen that both the reflection coefficient and the transmission coefficient decrease as the perturbation parameter $\varepsilon$ increases\rev{, within the \revB{energy-consistency range} of the expansion}. That is to say, due to the permeability effect, the total wave energy is dissipated while going through the permeable plate. Especially the decline of the wave energy becomes more evident as $a/b$ is smaller, which means that the permeability effect is stronger as the breakwater stretches further down from the water surface. \rev{The decrease of the transmission coefficient with increasing $\varepsilon$ may appear counter-intuitive, since a highly porous plate must eventually become transparent to the waves ($T \to 1$ as $\varepsilon \to \infty$); that regime, however, lies outside the one-sided expansion about the impermeable plate, and within the \revB{energy-consistency range} $\varepsilon \le \varepsilon_{\max}(kb)$ the porous through-flow acts primarily as an energy sink, so that both $R$ and $T$ are reduced while $1 - R^2 - T^2 = \varepsilon\mathcal{D}_0 > 0$ measures the dissipated energy flux (see \S5).}

\revB{Gayen \& Mondal [14] solved the same problem, a thin porous plate submerged in deep water, by a hypersingular integral equation valid for arbitrary porosity; their plate condition and energy identity have the same form as (2.15) and (5.1), so that their porous-effect parameter $G$ is our $\varepsilon$. They tabulated $|R|$ and $|T|$ for $a/b = 0.25$ at $G = 0$ and $G = 1$. Table 2 sets their values against the present ones. At $G = 0$ the two agree to within $0.1$--$4\%$. At $G = 1$, which lies well inside the energy-consistency bound ($\varepsilon_{\max} = 4.8$--$28$ for this geometry) but far above the load-consistency bound $\varepsilon_F$ of \S4.3, the truncated first-order solution is no longer quantitatively useful. It gives $R$ larger than $R_0$ where the full solution gives $R$ smaller, and it overestimates the dissipated flux $1 - R^2 - T^2$ by a factor of $1.3$ to $3.6$. The first-order rate $\mathcal{D}_0$ is the slope of $1 - R^2 - T^2$ at $\varepsilon = 0$, whereas the values of [14] at $G = 1$ give secant slopes; since by (5.1) the dissipated flux is $2k\varepsilon\int_L|\phi^+ - \phi^-|^2\,dy$, the smaller secant slopes mean that the potential jump across the plate at $G = 1$ is markedly smaller than the leading-order jump, a reduction that the coupling in (2.15) produces only at second order. Values at small $G$ are not given in [14], so the slope $\mathcal{D}_0$ itself cannot be tested against that work. Thus (5.6) is a necessary condition only (\S5.3), and for $a/b = 0.25$ the first-order theory should be used only for $\varepsilon$ well below $\varepsilon_F$.}

\begin{table}[htbp]
\centering
\revB{\caption{Comparison with the hypersingular-integral-equation solution of Gayen \& Mondal [14, tables 1 and 2] for $a/b = 0.25$ (their $d/p = 0.625$, $b/p = 0.375$; $Kp = kb$; $G = \varepsilon$). Columns marked [14] are their values; the others are the present leading-order ($\varepsilon = 0$) and truncated first-order ($\varepsilon = 1$) values, and $\mathcal{D}_0$ of (5.3) against the secant dissipation rate $(1 - R^2 - T^2)/G$ of [14] at $G = 1$.}
\scriptsize\setlength{\tabcolsep}{3.5pt}
\begin{tabular}{c|cc|cc|cc|cc|cc|cc}
\hline
 & \multicolumn{4}{c|}{$\varepsilon = G = 0$} & \multicolumn{6}{c|}{$\varepsilon = G = 1$} & & \\
$kb$ & $R_0$ & [14] & $T_0$ & [14] & $R$ & [14] & $T$ & [14] & $R^2\!+\!T^2$ & [14] & $\mathcal{D}_0$ & [14] \\
\hline
0.6 & 0.0848 & 0.0814 & 0.9964 & 0.9967 & 0.0891 & 0.0751 & 0.9610 & 0.9688 & 0.9314 & 0.9442 & 0.071 & 0.056 \\
1.5 & 0.1731 & 0.1664 & 0.9849 & 0.9861 & 0.2153 & 0.1136 & 0.8265 & 0.9220 & 0.7294 & 0.8630 & 0.323 & 0.137 \\
2.0 & 0.1710 & 0.1696 & 0.9853 & 0.9855 & 0.2379 & 0.1005 & 0.7917 & 0.9265 & 0.6834 & 0.8684 & 0.394 & 0.132 \\
2.5 & 0.1543 & 0.1494 & 0.9880 & 0.9888 & 0.2396 & 0.0838 & 0.7826 & 0.9369 & 0.6698 & 0.8848 & 0.417 & 0.115 \\
\hline
\end{tabular}}
\end{table}

\begin{figure}[htbp]
\centering
\includegraphics[width=0.95\textwidth]{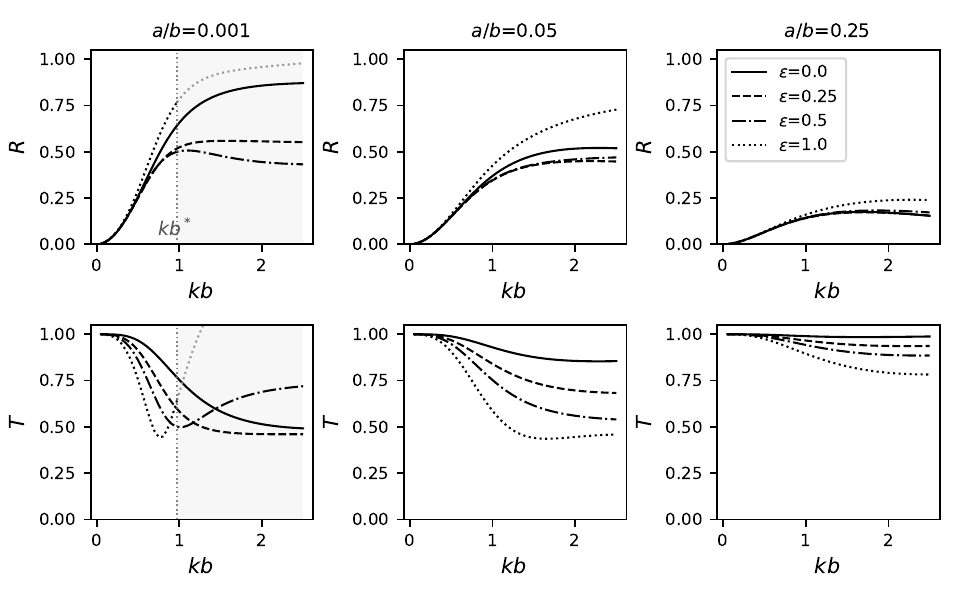}
\caption{(top) Reflection coefficient $R$ and (bottom) transmission coefficient $T$ versus $kb$ for (a) $a/b = 0.001$; (b) $a/b = 0.05$; (c) $a/b = 0.25$, for $\varepsilon = 0,\ 0.25,\ 0.5,\ 1$. \rev{Grey segments indicate $\varepsilon > \varepsilon_{\max}(kb)$, where the truncated expansion violates energy conservation; vertical lines mark $kb^*$ with $\varepsilon = \varepsilon_{\max}(kb^*)$. (The authors' original figure (2023), recomputed with scripts written with Anthropic Claude (Fable~5); results verified via the energy identity (5.4).)}}
\end{figure}

\revB{Figure 10 shows the velocity on the plate for $kb = 1$. Panel (a) gives the leading-order potential jump $|\phi_0^+ - \phi_0^-|$, which by (2.19) is also the amplitude of the first-order normal velocity through the plate, $|\partial\phi_1/\partial x| = k|\phi_0^+ - \phi_0^-|$. The seepage vanishes at both edges and is largest in the upper third of the plate, where the wave motion is strongest. Panel (b) gives the tangential velocity $|\partial\phi_0^+/\partial y|$ on the $+$ face against the distance $r$ from the nearer edge on logarithmic scales; the local slope is $-0.50$ at both edges for both values of $a/b$, i.e. the inverse-square-root singularity that follows from (3.8) (\S2.1), integrable and with finite energy. Because of this singularity the local velocity exceeds any fixed breaking criterion within a small neighbourhood of each edge, whose size shrinks with the wave steepness. The condition $kA \ll 1$ of \S2.1 is therefore a global criterion, and the local behaviour at the edges lies outside the scope of the linear theory, as for the impermeable plate [8]. Since $\phi_1$ satisfies the same Riemann--Hilbert structure, (4.15), the first-order velocity has the same behaviour at the edges.

Figure 11 shows the amplitude of the horizontal force (4.56) and the centre of pressure (4.59) against $kb$ for the same values of $\varepsilon$ and $a/b$ as in figure 9, curves beyond the energy-consistency bound being drawn in grey. For $\varepsilon = 0.25$ the force is below the impermeable value for $kb \ge 0.2$ ($a/b = 0.001$ and $0.05$) and for $0.25 \le kb \le 2.1$ ($a/b = 0.25$), the reduction growing with $kb$ and being largest for the plate extending furthest below the surface: at $kb = 2.5$ it is $36\%$ for $a/b = 0.001$, $13\%$ for $a/b = 0.05$ and about $1\%$ for $a/b = 0.25$. The centre of pressure changes by less than $0.05$ of the plate length over the range shown. For $\varepsilon = 0.5$ with $a/b = 0.25$, and for $\varepsilon = 1$ with all three geometries, the truncated amplitude exceeds the impermeable value even inside the energy-consistency bound. The increase is the quadratic term discussed in \S4.3 and is an artefact of the truncation. The load-consistency bound $\varepsilon_F$ of \S4.3 is $0.61$, $0.83$, $0.90$ for $a/b = 0.001$, $0.54$, $0.69$, $0.63$ for $a/b = 0.05$ and $0.38$, $0.42$, $0.27$ for $a/b = 0.25$ at $kb = 0.5$, $1$, $2$ respectively, and it tends to zero in the long-wave limit, where the correction becomes exactly in quadrature with the leading-order force. The force is thus a more demanding quantity than the scattering coefficients. Whereas $R$ and $T$ remain energy-consistent up to $\varepsilon_{\max}$, which exceeds unity for $a/b \ge 0.05$ over the whole range of $kb$, the first-order load should be used only for $\varepsilon$ well below $\varepsilon_F$, and a second-order theory is required beyond.}

\begin{figure}[htbp]
\centering
\includegraphics[width=0.95\textwidth]{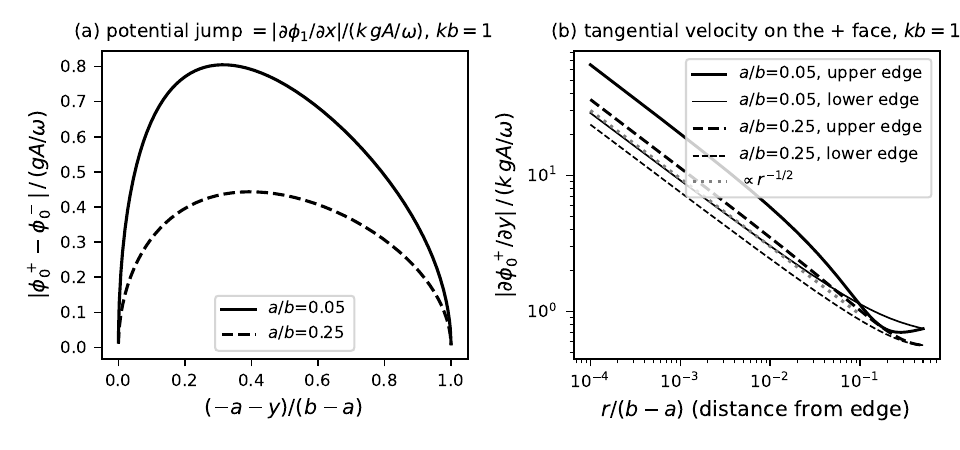}
\caption{\revB{Velocity on the plate at $kb = 1$ for $a/b = 0.05$ (---) and $0.25$ (- - -). (a) Leading-order potential jump $|\phi_0^+ - \phi_0^-|/(gA/\omega)$ along the plate, equal to the first-order normal velocity $|\partial\phi_1/\partial x|/(k\,gA/\omega)$ by (2.19). (b) Tangential velocity $|\partial\phi_0^+/\partial y|/(k\,gA/\omega)$ on the $+$ face against the distance from the upper (thick) and lower (thin) edge; the dotted line is $\propto r^{-1/2}$. (Script written with Anthropic Claude (Fable~5.1) under the authors' direction.)}}
\end{figure}

\begin{figure}[htbp]
\centering
\includegraphics[width=0.95\textwidth]{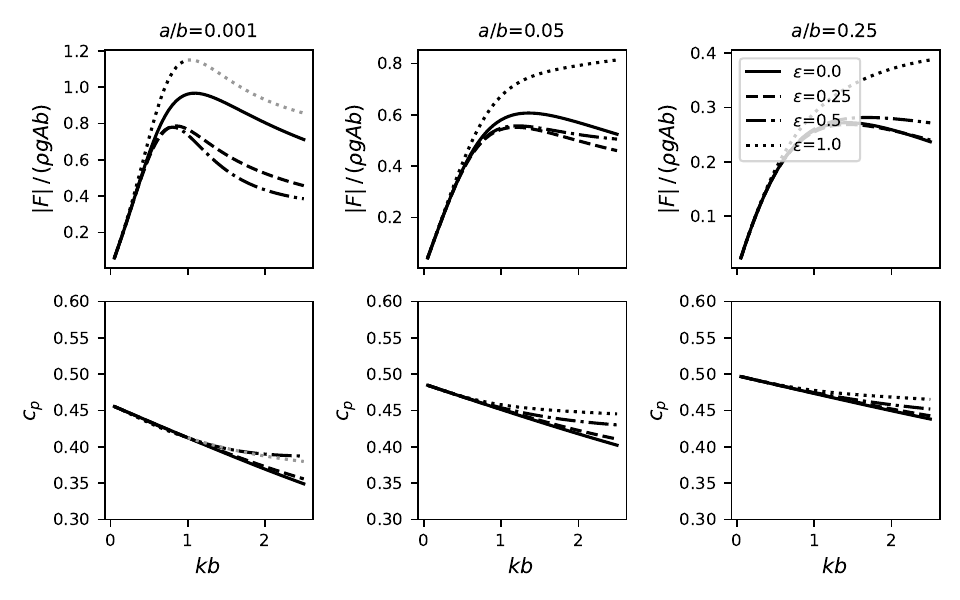}
\caption{\revB{(top) Amplitude of the horizontal force $|\mathfrak{F}|/(\rho g A b)$, (4.56), and (bottom) centre of pressure $c_p$, (4.59), versus $kb$ for (a) $a/b = 0.001$; (b) $a/b = 0.05$; (c) $a/b = 0.25$, for $\varepsilon = 0,\ 0.25,\ 0.5,\ 1$. Grey segments indicate $\varepsilon > \varepsilon_{\max}(kb)$. The $\varepsilon = 0$ curves coincide with Evans' equations (73) and (75) [8]. (Script written with Anthropic Claude (Fable~5.1) under the authors' direction; leading order verified against [8].)}}
\end{figure}

\section{Conclusion}
\revB{The main results of this study may be summarized as follows.
\begin{enumerate}
\item[(i)] \emph{Solution.} The two-dimensional scattering of linear monochromatic waves by a submerged vertical permeable plate of finite extent in deep water has been solved by a regular perturbation expansion in the porosity parameter $\varepsilon$. At each order the plate condition prescribes the normal velocity, and the method of reduced potentials converts the problem into a Riemann--Hilbert problem with a cut along the plate and its image: homogeneous at leading order, where the impermeable-plate solution is recovered, and nonhomogeneous at first order, where the forcing is the leading-order potential jump across the plate. In contrast to previous studies, which resolve the porous condition numerically over the whole range of the porous parameter, the expansion is about the impermeable-plate problem in the small-permeability limit, and the scattered field is obtained in closed form at each order. The leading-order reflection and transmission coefficients are obtained in closed form, and the far-field amplitudes at first order reduce to a single Cauchy principal-value integral, evaluated here by Chebyshev quadrature with subtraction of the singularity; no algebraic system has to be assembled or inverted.
\item[(ii)] \emph{Energy identity and consistency bound.} An exact energy identity (5.1) holds for the full permeable-plate problem at any porosity. Its first-order form (5.4) is satisfied identically by the derived solution, which provides an independent check of the analysis of \S4. For the truncated expansion, energy is conserved if and only if $\varepsilon \le \varepsilon_{\max}(kb) = \Rej\{\Lambda\Delta_{123}\}/|\Lambda|^2$, an energy-consistency bound obtained in closed form; within it the dissipated energy flux is $1 - R^2 - T^2 = \varepsilon\mathcal{D}_0$, and beyond it the apparent growth of the coefficients at large $kb$ is an artefact of the truncation.
\item[(iii)] \emph{Applicability.} The results hold for $kA \ll 1$, $A < a$ and real $\varepsilon$; the truncated expansion is energy-consistent for $\varepsilon \le \varepsilon_{\max}(kb)$ and quantitatively reliable only for $\varepsilon$ well below the load-consistency bound $\varepsilon_F$ (\S\S4.3, 6). For a fixed plate the criterion $\varepsilon \le \varepsilon_{\max}$ is satisfied for long waves and fails at the short-wave end. Within the energy-consistency range both $R$ and $T$ decrease with increasing permeability, and the dissipation is strongest for plates extending far below the surface (small $a/b$).
\item[(iv)] \emph{Limitations.} The water is infinitely deep; the through-flow is purely resistive, so that $\varepsilon$ is real; the expansion is one-sided about the impermeable plate and cannot reach the transparent limit of large porosity; and the truncation at first order is not accompanied by an error estimate, so that (5.6) is a necessary rather than a sufficient condition for the accuracy of the expansion; the comparison with the full solution of [14] at $\varepsilon = 1$ in \S6 shows the truncation error there to be large for $a/b = 0.25$. The hydrodynamic load is more demanding than the scattering coefficients in this respect: its first-order correction is nearly in quadrature with the leading-order force, and the truncated load is reliable only for $\varepsilon$ well below the load-consistency bound $\varepsilon_F$ of \S4.3, which is considerably stricter than (5.6).
\item[(v)] \emph{Further work.} The second-order problem, which would furnish an error estimate and extend the energy-consistency range; complex porosity parameters representing inertial effects in the pores; finite depth; and the surface-piercing limit $a/b \to 0$, which is physically distinct from the smallest ratio computed here ($a/b = 0.001$) because the plate must then intersect the instantaneous free surface, requiring a separate treatment of the upper-edge condition.
\end{enumerate}}

\revB{\appendix
\section{Derivation of the energy identity (5.1)}
Let $D$ be the fluid domain $y < 0$ with the plate $L$ removed, bounded by the free surface $y = 0$, the plate faces $L^\pm$, the vertical lines $x = \pm X$ with $X \to \infty$, and $y \to -\infty$. Since $\phi$ and its complex conjugate $\bar\phi$ (conjugation with respect to $j$) are both harmonic in $D$, Green's second identity gives
\begin{equation}
\oint_{\partial D} \left(\bar\phi\,\frac{\partial\phi}{\partial n} - \phi\,\frac{\partial\bar\phi}{\partial n}\right) ds = 2j\,\Imj\oint_{\partial D} \bar\phi\,\frac{\partial\phi}{\partial n}\, ds = 0, \tag{A.1}
\end{equation}
with $n$ the outward normal. On $y = 0$, $\partial\phi/\partial n = \partial\phi/\partial y = k\phi$ by (2.3), so $\bar\phi\,\partial\phi/\partial n = k|\phi|^2$ is real and contributes nothing to (A.1); the contribution from $y \to -\infty$ vanishes by (2.4). On $x = +X$, with (2.8) written as $\phi = e^{ky}(X^{+\infty}e^{jkx} + X_I e^{-jkx})$, $X_I = -jgA/\omega$,
\begin{equation}
\begin{split}
\Imj\int_{-\infty}^{0} \bar\phi\,\frac{\partial\phi}{\partial x}\, dy &= \Imj\int_{-\infty}^{0} jk\,e^{2ky}\left(|X^{+\infty}|^2 - |X_I|^2 + 2j\,\Imj\{\bar X_I X^{+\infty}e^{2jkx}\}\right) dy\\
&= \tfrac12\left(|X^{+\infty}|^2 - |X_I|^2\right),
\end{split} \tag{A.2}
\end{equation}
the cross term being real after multiplication by $jk$; on $x = -X$, where the outward normal is $-x$ and (2.9) holds, the contribution is $+\tfrac12|X^{-\infty}|^2$. On the plate faces the outward normals are $\mp x$ and $\partial\phi/\partial x$ is continuous across $L$ by (2.15), so that
\begin{equation}
\begin{split}
\Imj\left(-\int_L \bar\phi^+\frac{\partial\phi}{\partial x}\, dy + \int_L \bar\phi^-\frac{\partial\phi}{\partial x}\, dy\right) &= -\Imj\int_L \overline{(\phi^+ - \phi^-)}\left(-jk\varepsilon\right)(\phi^+ - \phi^-)\, dy\\
&= k\varepsilon\int_L |\phi^+ - \phi^-|^2\, dy,
\end{split} \tag{A.3}
\end{equation}
where (2.15)--(2.16) have been used in the form $\partial\phi/\partial x = -jk\varepsilon(\phi^+ - \phi^-)$. Collecting (A.2)--(A.3) in (A.1) and dividing by $|X_I|^2 = (gA/\omega)^2$ gives, with $R = |X^{+\infty}|/|X_I|$ and $T = |X^{-\infty}|/|X_I|$, the identity (5.1). If $r = X^{+\infty}/X_I = r_0 + \varepsilon r_1$, $t = X^{-\infty}/X_I = t_0 + \varepsilon t_1$ and $\phi^+ - \phi^- = (\phi_0^+ - \phi_0^-) + O(\varepsilon)$ are substituted in (5.1) and $|r_0|^2 + |t_0|^2 = 1$ is used, the terms of order $\varepsilon$ give (5.3); the leading-order jump is $\phi_0^+ - \phi_0^- = -2e^{ky}\xi_0(-k, y)$ by (4.8), which gives the expression for $\mathcal{D}_0$ in (5.3). Finally, with (5.2) and $\Delta_{123} = \Delta_{11} - j(\Delta_{12} - \Delta_{13})$,
\begin{equation}
\bar r_0 r_1 + \bar t_0 t_1 = -\frac{\Lambda\left(\Delta_{11} - j(\Delta_{12} - \Delta_{13})\right)}{|\Delta_{123}|^2} = -\frac{\Lambda\,\Delta_{123}}{|\Delta_{123}|^2}, \tag{A.4}
\end{equation}
so that (5.3) becomes (5.4).}

\section*{CRediT authorship contribution statement}
\textbf{Jeongin Kim:} formulated the problem, derived the closed-form solutions and carried out the original numerical implementation; \textbf{Yong Sung Park:} conceived and supervised the study, derived the energy identity and the \revB{energy-consistency bound}, verified the analysis, and revised the manuscript. Both authors gave final approval for publication.

\section*{Declaration of competing interest}
The authors declare that they have no known competing financial interests or personal relationships that could have appeared to influence the work reported in this paper.

\section*{Data availability}
The Python scripts that generate all figures and verify the energy identity (5.4), together with the plotted data, are archived at Zenodo: \url{https://doi.org/10.5281/zenodo.21818444} [40].

\section*{Declaration of generative AI and AI-assisted technologies in the writing process}
The original manuscript (2023) was written by the authors without AI assistance. During the preparation of the 2026 revision, the authors used Anthropic Claude (Fable~5) to draft the added section on the energy identity and the \revB{energy-consistency bound} (\S5) and to revise limited passages of the existing text\revB{, and, during the revision in response to the reviewers, Anthropic Claude (Fable~5.1) to draft the amended and added passages (\S\S1--2, 4.3, 6--7, Appendix~A) and the additional scripts (comparison with [8], plate loads, figures 3, 10 and 11) under the authors' direction, with all results verified by the authors as described in \S\S3.2, 4.3 and 6}. After using this tool, the authors reviewed and edited the content as needed and take full responsibility for the content of the publication. Uses of AI-assisted tools in the research process itself are disclosed as follows: the numerical scripts archived at Zenodo [40] were implemented, and supporting algebra was carried out, with Anthropic Claude (Fable~5) under the direction of the corresponding author, with all results verified independently by the authors (see \S6 and the figure captions); the literature was additionally surveyed with Anthropic Claude (Sonnet~5) and Google Gemini (3.1~Pro) through a batch interface, and every reference was checked against its published record.

\section*{Acknowledgements}
\revB{This research was supported by the National Research Foundation of Korea (NRF) grant funded by the Korea government (MSIT) \mbox{[RS-2026-25475182]}. The authors acknowledge support from the Institute of Engineering Research, the Institute of Construction and Environmental Engineering and the Institute for Peace and Unification Studies at Seoul National University.}

\end{document}